\documentclass[aps,prd,twocolumn,superscriptaddress,floatfix,nofootinbib]{revtex4-2}

\usepackage{amsmath,amssymb,mathtools,bm}
\usepackage{graphicx}
\usepackage{hyperref}
\usepackage{microtype}
\usepackage{mathrsfs}

\hypersetup{colorlinks=true,linkcolor=blue,citecolor=blue,urlcolor=blue}
\allowdisplaybreaks[4]

\newcommand{\id}{\mathbf 1}
\newcommand{\cV}{\mathcal V}
\newcommand{\cBalpha}{\mathcal H_{\mathrm B,\alpha}}
\newcommand{\cBleft}{\mathcal H_{\mathrm B,\ell}}
\newcommand{\cBright}{\mathcal H_{\mathrm B,\mathrm r}}
\newcommand{\Tr}{\operatorname{Tr}}
\newcommand{\Spec}{\operatorname{Spec}}

\newcommand{\dd}{\mathrm d}
\newcommand{\PP}{\mathsf P}
\newcommand{\Q}{\mathsf Q}
\newcommand{\sL}{\mathsf L}
\newcommand{\rr}{\mathrm r}

\begin{document}
	
	\title{Time-dependent integrable quantum field theories with dynamical boundaries}
	
	\author{Pradip Kattel}
	\email{pradip.kattel@unige.ch}
	\affiliation{Department of Quantum Matter Physics, University of Geneva,
		Quai Ernest-Ansermet 24, 1211 Geneva, Switzerland}
	
	\begin{abstract}
		We present a general method to construct and solve time-dependent integrable chiral quantum field theories on finite intervals from autonomous scattering and reflection data. Spatial homogeneity forces the spectral labels of right- and left-moving particles to be affine with a common slope, so bulk and boundary couplings sample one spectral trajectory. A complete round trip shifts the spectral labels and gives boundary quantum Knizhnik--Zamolodchikov equations. The Yang--Baxter and reflection equations provide local consistency, while boundary-qKZ flatness provides global consistency. We reconstruct the local Hamiltonian by Cayley inversion of the bulk and boundary matching matrices. We apply the construction to rational $SU(N)$ scattering with fundamental boundary impurities and to an $O(3)$ model with a spin-$\tfrac12$ impurity. For the $SU(2)$ impurity, we construct explicit boundary-qKZ solutions as lattice sums of off-shell boundary Bethe vectors and recover the static open-chain Bethe equations when the drive rate vanishes.\end{abstract}
	\maketitle
	
	\section{Introduction}
	\label{sec:introduction}
	
	An integrable quantum field theory has an infinite number of conserved charges. In $1+1$ dimensions, these conservation laws preclude particle production during scattering events. The scattering particles might exchange their momenta, but no new momenta are produced. This leads to factorized scattering, in which the multiparticle $S$-matrix factorizes into products of two-particle $S$-matrices. Consistency of different factorizations of three-particle scattering gives the Yang--Baxter equation~\cite{ZZ1979,Faddeev1996}. On a finite spatial interval of length $\sL$, particles also reach the endpoints and reflect. A reflection matrix describes each endpoint process. Comparing the two possible orders of a two-particle reflection process gives the reflection equation~\cite{Cherednik1984,Sklyanin1988,GhoshalZamolodchikov1994}. 
	
	Many integrable quantum field theories with boundaries have been studied, including the sine-Gordon model, affine Toda field theories, nonlinear sigma models, and Gross--Neveu models~\cite{GhoshalZamolodchikov1994,FringKoberleToda1994,
		GhoshalON1994,DeMartinoMoriconi1999}. Their exact solutions give boundary spectra and bound states, finite-size and thermodynamic properties, and access to nontrivial boundary renormalization group flows~\cite{LeClairMussardoSaleurSkorik1995,DoreyPocklingtonTateoWatts1998,
		DoreyRunkelTateoWatts2000,rylands2020exact}. Integrable boundary methods have also been used to study quantum impurities in interacting media, including superconducting wires, interacting spin chains, and other strongly correlated hosts~\cite{PasnooriRylandsAndrei2020,PasnooriYSR2022,KattelXXZ2025,kattel2026multichannel}.
	
	The bulk and boundary interactions considered here depend on time. Integrability at each fixed time does not by itself give a conserved set of charges for the driven evolution. Scattering and reflection events separated in time have to be compared with their spectral arguments included. On a finite interval this comparison is made by following a particle through successive reflections and bulk exchanges. Free propagation together with spatial homogeneity fixes the labels carried by the particles to be affine functions of their characteristic coordinates. All local events then sample the autonomous data along the single trajectory $u_B(t)=u_{B,0}+\kappa t$. This characteristic geometry fixes the time dependence of the bulk and boundary couplings.
	
	The additional consistency requirement is a form of path independence. A familiar example appears in the zero-curvature formulation of integrable partial differential equations. Consider a field $q(x,t)$ satisfying a nonlinear equation of motion. For an integrable partial differential equation, an auxiliary vector $\Phi(x,t;\lambda)$ can be introduced that satisfies
	\begin{align}
		\left(\partial_x-A_x[q;\lambda]\right)\Phi&=0,
		\label{eq:introauxx}\\
		\left(\partial_t-A_t[q;\lambda]\right)\Phi&=0,
		\label{eq:introauxt}
	\end{align}
	where $\lambda$ is a spectral parameter and the matrices $A_x$ and $A_t$ are constructed from $q$ and its derivatives. These equations describe the evolution of one auxiliary vector in the $x$ and $t$ directions, so they must be compatible. Writing $\nabla_x=\partial_x-A_x$ and $\nabla_t=\partial_t-A_t$, compatibility requires $[\nabla_t,\nabla_x]\Phi=0$, which gives
	\begin{equation}
		\partial_tA_x-\partial_xA_t+[A_x,A_t]=0.
		\label{eq:introzerocurvature}
	\end{equation}
	The zero-curvature condition encodes the nonlinear dynamics as compatibility of the two auxiliary evolutions. Geometrically, it states that the associated transport is locally path independent. Related compatibility constructions for time-dependent quantum Hamiltonians lead to Knizhnik--Zamolodchikov equations in parameter space~\cite{SinitsynYuzbashyan2018,Yuzbashyan2018}.
	
	A similar compatibility condition appears here in a discrete form. Particle ordering and chirality divide configuration space into sectors. Bulk scattering relates neighboring ordering sectors, and reflection relates incoming and outgoing chiral sectors. The Yang--Baxter and reflection equations make these local relations compatible. A particle can also reflect at both endpoints and return after crossing the other particles. The corresponding events occur at different times, so their spectral arguments change along the path. The two possible orders of complete scattering and reflection circuits must nevertheless give an identical amplitude~\cite{KattelGeneral2026}.
	
	The characteristic map fixes the spectral arguments along each trajectory. A particle carries its label unchanged during free motion, and the autonomous matrices are evaluated on the labels brought together at a scattering or reflection event. The spacetime wavefunction is obtained by pulling the resulting coefficient function back along these characteristic labels. The use of characteristic labels to pull autonomous factorized data back to a time-dependent chiral field theory was introduced in Ref.~\cite{KattelGeneral2026}, whose framework we follow here.
	
	Given autonomous bulk $S$-matrix and boundary $K$-matrices, the construction determines the allowed time dependence of the bulk and boundary interactions and reduces the exact finite-particle dynamics to a boundary qKZ problem. The bulk and boundary couplings are not introduced as independent functions of time. Their time dependence follows from the characteristic map and the consistency of the resulting factorized transport.
	
	Spatial homogeneity first fixes the characteristic map. Bulk scattering must obey one local rule at every position. Writing
	\begin{equation}
		\zeta_R=f_R(x-t), \qquad \zeta_L=f_L(x+t),
		\label{eq:intro-characteristic-labels}
	\end{equation}
	the difference of the two labels at a right--left scattering event is $f_L(x+t)-f_R(x-t)$. It can depend on the scattering time, but spatial homogeneity requires it to be independent of the scattering position. Thus its derivative with respect to $x$ vanishes, and since $x+t$ and $x-t$ are independent variables, both functions must have the same constant derivative. Thus,
	\begin{equation}
		f_R(\xi)=\kappa\xi+\nu_R, \qquad
		f_L(\xi)=\kappa\xi+\nu_L,
		\label{eq:intro-affine-labels}
	\end{equation}
	and hence the difference at a scattering event is $2\kappa t+\beta$ with $\beta=\nu_L-\nu_R$. The bulk matrix sampled at time $t$ is $S(2\kappa t+\beta)$. The common slope $\kappa$ sets the drive rate, while $\kappa\to 0$ gives the autonomous limit.
	
	This affine map also determines the spectral geometry at the two endpoints. At each endpoint, the sum of the incoming and outgoing labels is independent of time, so reflection occurs about a fixed center in spectral space. Because the two endpoint centers differ, two successive reflections translate the spectral label by $\Delta=2\kappa\sL$. A complete round trip is represented by a difference transport in spectral space, and the associated consistency equations are the boundary quantum Knizhnik--Zamolodchikov equations~\cite{FrenkelReshetikhin1992,Cherednik1992,PasnooriBoundaryqKZ2026}. Requiring different complete scattering and reflection circuits to produce the same amplitude gives the BqKZ flatness relation. This relation is the global consistency condition that supplements the local Yang--Baxter and reflection equations. The construction is formulated directly as a boundary transport problem on the finite interval.
	
	\section{Summary of the construction}
	\label{sec:summary}
	
	Let $S(u)$ be a unitary bulk scattering matrix and let $K_\ell(u)$ and $K_\rr(u)$ satisfy the corresponding reflection equations. Assign a spectral label to each chiral particle. Requiring the difference of the labels at a fixed-time scattering event to be independent of position gives
	\begin{equation}
		\zeta_R=\kappa(x-t)+\nu_R,\qquad \zeta_L=\kappa(x+t)+\nu_L.
	\end{equation}
	Consequently, all bulk scattering at time $t$ uses one argument $2u_B(t)$, with $u_B(t)=u_{B,0}+\kappa t$, while the two endpoints use $u_B(t)$ and $-u_B(t)$.
	
	Following a particle from one endpoint to the other and back shifts its spectral label by $\Delta=2\kappa\sL$. The consistency of different sequences of such round trips is the boundary qKZ system. This is the global condition in addition to the local Yang--Baxter and reflection equations. A solution of the boundary qKZ equations is pulled back through the affine characteristics to obtain the time-dependent many-body wavefunction.
	
	These data fix the local Hamiltonian. The two-particle matching matrix gives the bulk contact interaction through a Cayley inverse, while the one-particle matching matrix gives the endpoint interaction. The explicit realizations below use rational $SU(N)$ data and an $O(3)$ model. For the $SU(2)$ impurity, the impurity is treated as an additional site with fixed spectral label, and the coefficient function is constructed from off-shell boundary Bethe vectors summed over shifted root lattices. The nested $SU(N)$ case uses the corresponding nested Bethe vectors.
	
	\section{Setup and spectral data}
	\label{sec:conventions}
	
	\subsection{The interval and its fields}
	
	We define the theory on the spatial interval $0\leq x\leq\sL$, with the left and right endpoints at $x_\ell=0$ and $x_\rr=\sL$, respectively. Each particle carries an internal degree of freedom in the one-particle space $\cV$, with $N=\dim\cV$. We denote the right- and left-moving fields by $\psi_{R,a}(x)$ and $\psi_{L,a}(x)$, where $a=1,\ldots,N$, and repeated internal indices are summed. The free chiral Hamiltonian is
	\begin{equation}
		H_0=-i\int_0^{\sL}\dd x\,\psi_{R,a}^\dagger\partial_x\psi_{R,a}+i\int_0^{\sL}\dd x\,\psi_{L,a}^\dagger\partial_x\psi_{L,a},
		\label{eq:H0}
	\end{equation}
	so that right and left movers have velocities $+1$ and $-1$. We work with these bare chiral fields and their Fock vacuum $|0\rangle$. The many-body construction is carried out in the finite-particle-number sectors generated by $\psi_{R,a}^\dagger$ and $\psi_{L,a}^\dagger$ above this vacuum. Boundary quantum degrees of freedom are kept separately in the endpoint Hilbert spaces introduced below. A particle reaches the left endpoint as a left mover and the right endpoint as a right mover, and reflection changes its chirality.
	
	In the bulk, a right mover and a left mover scatter when they meet. Denoting their positions by $x_R$ and $x_L$, we take $y=x_R-x_L$ as the relative coordinate. Since their velocities are $+1$ and $-1$, $\dot y=2$. Before scattering $x_R<x_L$, so $y<0$, while after scattering $y>0$. We call these the incoming and outgoing regions and define $S$ as the matching matrix from incoming to outgoing. We write $\PP_{12}$ for the permutation on $\cV\otimes\cV$ and $S_{21}(u)=\PP_{12}S_{12}(u)\PP_{12}$. The bulk matrices used below are permutation symmetric, $S_{21}(u)=S_{12}(u)$.
	
	\subsection{Boundary degrees of freedom}
	\label{sec:boundaryspaces}
	
	When a particle reaches an endpoint and gets reflected, its chirality changes and its internal state can change as well. If the endpoint has no quantum degrees of freedom of its own, the reflection is described by an $N\times N$ matrix acting on $\cV$. Instead, if a quantum impurity is present at the end, the impurity adds another degree of freedom at the endpoint. We denote its Hilbert space by $\cBalpha$. The reflection matrix then acts on the joint particle--impurity space $\cV_a\otimes\cBalpha$, and the impurity state can change during reflection.
	
	We use a common notation for both cases by taking $\cBalpha=\mathbb C$ when no impurity is present. This is to indicate that there is no additional boundary quantum degree of freedom. The reflection matrix can still be a nontrivial $N\times N$ matrix on $\cV$. The two endpoints can be chosen independently, with spaces $\cBleft$ and $\cBright$. Thus the bulk particles are created above the bare Fock vacuum, while the boundary quantum states are carried by $\cBleft$ and $\cBright$. For $n$ bulk particles, the amplitudes take values in
	\begin{equation}
		\mathcal F_n=\cV^{\otimes n}\otimes\cBleft\otimes\cBright.
		\label{eq:coefficientspace}
	\end{equation}
	
	\subsection{Spectral labels and the autonomous data}
	\label{sec:autonomousdata}
	
	We start from the scattering and reflection data of an autonomous integrable theory. We assign a spectral label $\zeta_j$ to each particle. The bulk scattering matrix is of difference form. Thus particles carrying labels $\zeta_i$ and $\zeta_j$ scatter with $S(\zeta_i-\zeta_j)$. This is the scattering family used in the autonomous problem. The driven theory arises from how the moving particles carry these labels.
	
	Reflection families $K_\ell$ and $K_\rr$ describe the two endpoints. The particle labels also fix their physical arguments. Before we can determine them, we have to find how a label is carried by a freely moving right or left mover.
	
	\section{From free motion to time-dependent scattering}
	\label{sec:affine}
	
	\subsection{Labels carried by the chiral particles}
	
	The free Hamiltonian determines how we attach spectral labels to the particles. Eq.~\eqref{eq:H0} gives the characteristic derivatives $D_R=\partial_t+\partial_x$ and $D_L=\partial_t-\partial_x$. A right mover follows $x-t=\text{const}$, while a left mover follows $x+t=\text{const}$. We take the spectral label of a freely moving particle to remain constant along its trajectory. Thus, $D_R\zeta_R=0$ and $D_L\zeta_L=0$, which gives
	\begin{equation}
		\zeta_R=f_R(x-t),\qquad \zeta_L=f_L(x+t).
		\label{eq:characteristic-general}
	\end{equation}
	The functions $f_R$ and $f_L$ are not arbitrary once the interaction is required to be spatially homogeneous. Suppose a right mover and a left mover scatter at $(x,t)$. The matrix at that event is $S(\zeta_L-\zeta_R)$, with $\zeta_L-\zeta_R=f_L(x+t)-f_R(x-t)$. At a fixed time, the local interaction must be independent of where the particles meet. We require
	\begin{equation}
		\partial_x\bigl[\zeta_L(x,t)-\zeta_R(x,t)\bigr]=0.
		\label{eq:homog}
	\end{equation}
	This gives $f_L'(x+t)=f_R'(x-t)$. Since $x+t$ and $x-t$ are independent, both derivatives must equal the same constant $\kappa$. Hence
	\begin{equation}
		\zeta_R=\kappa(x-t)+\nu_R,\qquad \zeta_L=\kappa(x+t)+\nu_L.
		\label{eq:affine}
	\end{equation}
	The map from a free trajectory to its spectral label is affine, with a common slope for both chiralities. The constant $\kappa$ sets the drive rate. When $\kappa=0$, the labels no longer depend on the characteristic coordinate and the autonomous limit is recovered.
	
	\subsection{The bulk scattering argument}
	
	When a right mover and a left mover meet, their positions are equal. Eq.~\eqref{eq:affine} then gives $\zeta_L-\zeta_R=2\kappa t+\nu_L-\nu_R$. We write this difference as $2u_B(t)$, where
	\begin{equation}
		u_B(t)=u_{B,0}+\kappa t,\qquad u_{B,0}=\tfrac12(\nu_L-\nu_R).
		\label{eq:uB}
	\end{equation}
	Thus every right--left scattering event at time $t$ is described by $S(2u_B(t))$. We do not make the scattering family itself time dependent. The time dependence comes from the labels carried by the particles and from the value of their difference when they meet.
	
	At fixed $t$, the map $x_j\mapsto\zeta_j$ is invertible for $\kappa\neq0$. At $\kappa=0$, we understand the formulas below through their autonomous limit. If the bulk scattering is trivial, spatial homogeneity alone no longer fixes the characteristic map to the affine form given by Eq.~\eqref{eq:affine}. The boundary drive is nevertheless constrained by the consistency of the resulting many-body transport, as in the time-dependent Kondo construction of Ref.~\cite{PasnooriKondo2025}.
	
	\subsection{Reflection and the endpoint spectral coordinates}
	
	The affine labels also determine the arguments of the reflection matrices. At the endpoint $x=x_\alpha$, Eq.~\eqref{eq:affine} gives
	\begin{equation}
		\zeta_R+\zeta_L=2m_\alpha,\qquad m_\alpha=\kappa x_\alpha+\tfrac12(\nu_R+\nu_L).
		\label{eq:center}
	\end{equation}
	Thus reflection sends an incoming label $\zeta$ to $2m_\alpha-\zeta$. Measuring the label from the reflection center, $u_\alpha=\zeta-m_\alpha$, reduces this to $u_\alpha\to-u_\alpha$.
	
	At the left endpoint, a particle arrives as a left mover, and Eq.~\eqref{eq:affine} gives $u_\ell(t)=u_B(t)$. At the right endpoint, a particle arrives as a right mover and $u_\rr(t)=-u_B(t)$. We evaluate the physical reflection matrices as $K_\ell(u_B(t))$ and $K_\rr(-u_B(t))$.
	
	The two reflection centers differ by $m_\rr-m_\ell=\kappa\sL$. In the common coordinate $z=\zeta-m_\ell$, successive reflections give
	\begin{equation}
		z\stackrel{\ell}{\longmapsto}-z\stackrel{\rr}{\longmapsto}z+\Delta,\qquad \Delta=2\kappa\sL.
		\label{eq:translation}
	\end{equation}
	A complete round trip shifts the spectral label by $\Delta$, which is the finite shift appearing in the boundary qKZ equations. Since the right reflection center lies at $z=\Delta/2$, its endpoint-centered argument is $z-\Delta/2$.
	
	\subsection{Same-chirality exchange}
	\label{sec:chiralexchange}
	
	The Hamiltonian contains a physical bulk interaction only when a right mover and a left mover meet. Particles of the same chirality have the same velocity, so their relative coordinate is constant,
	\begin{equation}
		\frac{\dd}{\dd t}(x_j-x_k)=0\qquad\text{for }\chi_j=\chi_k,
		\label{eq:frozenrelative}
	\end{equation}
	and they never scatter dynamically. Same-chirality exchange enters only in relating different ordering sectors of the factorized wavefunction.
	
	For these auxiliary exchanges we use the difference-form family
	\begin{equation}
		\mathcal S^{RR}_{jk}=\mathcal S^{LL}_{jk}=S_{jk}(\zeta_j-\zeta_k),\qquad \mathcal S^{RL}_{jk}=S_{jk}(\zeta_L-\zeta_R).
		\label{eq:allexchange}
	\end{equation}
	This is an auxiliary continuation of the bulk $S$-matrix to the same-chirality ordering sectors, not an additional physical scattering process. It makes the three-particle consistency conditions instances of one Yang--Baxter equation and introduces no additional local interaction. In the autonomous limit the labels within each chirality sector coincide, so the auxiliary same-chirality exchange reduces to $S(0)$. For the families used below,
	\begin{equation}
		S(0)\propto\PP.
		\label{eq:SzeroPermutation}
	\end{equation}
	Thus, up to a scalar phase, the auxiliary same-chirality exchange reduces to the permutation convention of the coordinate Bethe ansatz~\cite{andrei1979diagonalization,AndreiFuruyaLowenstein1983}.
	
	\subsection{Local compatibility of scattering and reflection}
	\label{sec:factorization}
	
	The characteristic geometry assigns the spectral argument to each scattering and reflection event. Local factorization requires two sequences with the same initial and final configuration to give the same amplitude. In the bulk, three particles can be taken between the same initial and final ordering by two different sequences of pairwise exchanges. Equating these two sequences gives the Yang--Baxter equation
	\begin{equation}
		\begin{split}
			&S_{12}(u-v)S_{13}(u-w)S_{23}(v-w)\\
			&\qquad=S_{23}(v-w)S_{13}(u-w)S_{12}(u-v).
		\end{split}
		\label{eq:YBE}
	\end{equation}
	In addition, we require braiding unitarity,
	\begin{equation}
		S_{12}(u)S_{21}(-u)=\id.
		\label{eq:Sunitarity}
	\end{equation}
	This ensures that an exchange followed by the reverse exchange returns the original amplitude. For the unitary models considered below, the scattering matrices also satisfy $S_{12}^\dagger(u)S_{12}(u)=\id$ for the real spectral arguments describing the physical scattering processes.
	
	At the left endpoint, take two particles with incoming centered labels $u$ and $v$. The first exchange carries the difference $u-v$. Once one particle has reflected, its label changes sign, so the crossing between the two reflections carries the sum $u+v$. If the two reflections occur at times $t_1$ and $t_2$, that crossing occurs at $\bar t=(t_1+t_2)/2$; since $u_B(t)$ is affine, $u_B(t_1)+u_B(t_2)=2u_B(\bar t)$, exactly the physical bulk argument at the crossing. Requiring the two possible orders of these events to agree gives
	\begin{equation}
		\begin{split}
			&S_{12}(u-v)K_{1\ell}(u)S_{21}(u+v)K_{2\ell}(v)\\
			&\qquad=K_{2\ell}(v)S_{12}(u+v)K_{1\ell}(u)S_{21}(u-v).
		\end{split}
		\label{eq:RE}
	\end{equation}
	We also require reflection unitarity, $K_\alpha(u)K_\alpha(-u)=\id$, so that reversing the spectral reflection returns the original state. For the unitary models considered below, the reflection matrices also satisfy $K_\alpha^\dagger(u)K_\alpha(u)=\id$ for the real spectral arguments describing physical reflections.
	
	The right endpoint has the opposite orientation. Particles arrive there as right movers and leave as left movers, so the same boundary process is read in the opposite sense. The corresponding relation is
	\begin{equation}
		\begin{split}
			&S_{12}(v-u)K_{1\rr}(u)S_{21}\bigl(-(u+v)\bigr)K_{2\rr}(v)\\
			&\qquad=K_{2\rr}(v)S_{12}\bigl(-(u+v)\bigr)K_{1\rr}(u)S_{21}(v-u).
		\end{split}
		\label{eq:REright}
	\end{equation}
	The signs reflect the orientation of the right-centered spectral coordinate. In the right-centered variables $\bar u=-u$ and $\bar v=-v$, Eq.~\eqref{eq:REright} has the corresponding algebraic form of Eq.~\eqref{eq:RE}. Equivalently, in the present convention, if $\widetilde K(u)$ solves the left reflection equation, then $K_\rr(u)=\widetilde K(-u)$ gives the corresponding right-boundary family. A nontrivial impurity placed at the right endpoint must be written in this reflected form.
	
	The Yang--Baxter and reflection equations make the local exchanges and reflections path independent. On a finite interval, a particle can also travel from one end to the other and return. This round trip changes its label by $\Delta$ and leads to the global transport considered next.
	
	\section{Round-trip transport and the boundary qKZ system}
	\label{sec:transport}
	
	\subsection{Amplitudes in spectral space}
	
	For particle $j$ of chirality $\chi_j=\pm1$, use
	\[
	\zeta_j(x_j,t)=\kappa(x_j-\chi_j t)+\nu_{\chi_j},
	\qquad \nu_{+1}=\nu_R,\quad \nu_{-1}=\nu_L .
	\]
	At fixed particle number $n$, the amplitudes take values in the space $\mathcal F_n$ of Eq.~\eqref{eq:coefficientspace}. Let $Q\in S_n$ label the ordering sector $\mathcal C_Q$, with $x_{Q_1}<\cdots<x_{Q_n}$, and write its spectral-space amplitude as $F_Q(\boldsymbol\zeta)$. The spacetime wavefunction in that sector is
	\begin{equation}
		\Psi_Q(\boldsymbol x,t)=F_Q\bigl(\zeta_1(x_1,t),\ldots,\zeta_n(x_n,t)\bigr).
		\label{eq:pullback}
	\end{equation}
	For $\kappa\neq0$, the map $\boldsymbol x\mapsto\boldsymbol\zeta$ is invertible at fixed $t$. We solve the factorized problem in spectral space and then use Eq.~\eqref{eq:pullback} to obtain the spacetime wavefunction.
	
	The full state in the $n$-particle sector is
	\begin{equation}
		|\Psi_n(t)\rangle=\sum_{Q\in S_n}\int_{\mathcal C_Q}\dd^n x\;F_Q\bigl(\boldsymbol\zeta(\boldsymbol x,t)\bigr)|\boldsymbol x,Q\rangle.
		\label{eq:boundaryfixedNstate}
	\end{equation}
	Here $|\boldsymbol x,Q\rangle$ is the bare Fock state obtained by applying the appropriate right- and left-moving creation operators in the order specified by $Q$. The particle indices and the two boundary quantum states are carried by $F_Q\in\mathcal F_n$.
	
	Across a right--left scattering surface,
	\begin{equation}
		F_{\rm out}(\boldsymbol\zeta)=S\bigl(\zeta_L-\zeta_R\bigr)F_{\rm in}(\boldsymbol\zeta),
		\label{eq:bulkEvent}
	\end{equation}
	while reflection at endpoint $\alpha$ gives
	\begin{equation}
		F_{\rm out}(\boldsymbol\zeta)=K_\alpha(u_\alpha)F_{\rm in}(\boldsymbol\zeta).
		\label{eq:boundaryEvent}
	\end{equation}
	On the physical scattering surface $x_R=x_L$, Eq.~\eqref{eq:bulkEvent} reduces to $F_{\rm out}=S(2u_B(t_c))F_{\rm in}$, because $\zeta_L-\zeta_R=2u_B(t_c)$ at the scattering time $t_c$. At the right endpoint, the reflection argument is measured from the right spectral center, displaced by $\Delta/2$ from the left one.
	
	Choose a reference ordering $Q_0$ and let $s_a$ exchange two adjacent particles. If $Q=Q_0s_{a_1}\cdots s_{a_m}$, repeated use of the exchange relation gives
	\begin{equation}
		F_Q(\boldsymbol\zeta)=\mathcal U_Q(\boldsymbol\zeta)F_{Q_0}(\boldsymbol\zeta),\qquad \mathcal U_Q=S_{a_m}\cdots S_{a_1},
		\label{eq:boundarysectortransport}
	\end{equation}
	where each factor is evaluated on the spectral-label difference of the two particles exchanged at that step. The Yang--Baxter equation makes the result independent of the sequence of adjacent exchanges used to reach $Q$, while the reflection equation gives the corresponding consistency when the path reaches an endpoint.
	
	\subsection{Round-trip transport}
	\label{sec:doublerow}
	
	The local relations determine amplitudes between neighboring sectors. A further relation is obtained by following particle $j$ through the interval and back. The particle reflects at both endpoints, crosses the other particles on each leg, and returns with its label shifted by $\Delta$. The ordered product along this path is denoted by $\mathcal T_j$.
	
	The factors in $\mathcal T_j$ are evaluated on the labels incident on each event. A right--left crossing uses the label of the left mover minus that of the right mover; an auxiliary same-chirality crossing uses the transported label minus the stationary label. The boundary argument is measured from the corresponding reflection center. In the common coordinate the left endpoint uses $z_j$, while the right endpoint uses $-z_j-\Delta/2$.
	
	Use the common coordinate of Sec.~\ref{sec:affine},
	\[
	z_j=\zeta_j-m_\ell,
	\qquad \Delta=2\kappa\sL,
	\]
	and order the labels so that particle $1$ is leftmost. In the reference configuration all particles are left movers. The round trip below gives the shift in Eq.~\eqref{eq:translation}; each factor corresponds to one exchange or reflection along the path.
	
	Particle $j$ first travels left, meeting particles $j-1,\ldots,1$. These are left movers, so each crossing is a same-chirality exchange with argument $z_j-z_k$ and, with the first-encountered factor rightmost,
	\begin{equation}
		\mathcal A_j=S_{j,1}(z_j-z_1)\,S_{j,2}(z_j-z_2)\cdots S_{j,j-1}(z_j-z_{j-1}).
		\label{eq:Aj}
	\end{equation}
	It reflects at the left endpoint, contributing $K_{\ell,j}(z_j)$ and becoming a right mover with label $-z_j$. Traveling right it crosses every other particle, each a left mover with label $z_k$, so each crossing is a genuine right--left event with argument $z_k-(-z_j)=z_j+z_k$, met in the order $1,\ldots,j-1,j+1,\ldots,n$:
	\begin{equation}
		\begin{split}
			\mathcal B_j={}&S_{j,n}(z_j+z_n)\cdots S_{j,j+1}(z_j+z_{j+1})\\
			&\times S_{j,j-1}(z_j+z_{j-1})\cdots S_{j,1}(z_j+z_1).
		\end{split}
		\label{eq:Bj}
	\end{equation}
	At the right endpoint the incoming label is $-z_j$ in the common left-centered coordinate. The right spectral center lies a distance $\Delta/2$ away, so the reflection contributes
	\begin{equation}
		K_{\rr,j}\left(-z_j-\tfrac{\Delta}{2}\right),
		\label{eq:Krfactor}
	\end{equation}
	and the outgoing label is $z_j+\Delta$. Particle $j$ returns leftwards, meeting $n,\ldots,j+1$, still left movers, so these are same-chirality exchanges with the shifted label:
	\begin{equation}
		\mathcal C_j=S_{j,j+1}(z_j+\Delta-z_{j+1})\cdots S_{j,n}(z_j+\Delta-z_n).
		\label{eq:Cj}
	\end{equation}
	The shift appears only in $\mathcal C_j$, because $\mathcal C_j$ describes the return of the transported particle after its label has changed from $z_j$ to $z_j+\Delta$.
	
	Since operators act to the left of the state, the transport operator is the ordered product
	\begin{equation}
		\mathcal T_j(\boldsymbol z)= \mathcal C_j\; K_{\rr,j}\left(-z_j-\tfrac{\Delta}{2}\right)\; \mathcal B_j\; K_{\ell,j}(z_j)\; \mathcal A_j ,
		\label{eq:doubleRow}
	\end{equation}
	recording, from the right, the outward reorderings, the left reflection, the intervening bulk crossings, the right reflection and the return reorderings. Its net effect on the label of particle $j$ is $z_j\mapsto z_j+\Delta$. This is the round-trip transport generated by the bulk and endpoint exchange relations.
	
	Two special cases make the structure explicit. For $n=1$ all exchange products are empty,
	\begin{equation}
		\mathcal T_1(z)=K_\rr\left(-z-\tfrac{\Delta}{2}\right)\,K_\ell(z),
		\label{eq:Tn1}
	\end{equation}
	so a single particle advances by $\Delta$ per round trip. For $n=2$, using the symmetry $S_{21}(u)=S_{12}(u)$,
	\begin{equation}
		\begin{split}
			\mathcal T_1={}&S_{12}(z_1-z_2+\Delta)\,K_{\rr,1}\left(-z_1-\tfrac{\Delta}{2}\right)\\
			&\times S_{12}(z_1+z_2)\,K_{\ell,1}(z_1),
		\end{split}
		\label{eq:T1n2}
	\end{equation}
	\begin{equation}
		\begin{split}
			\mathcal T_2={}&K_{\rr,2}\left(-z_2-\tfrac{\Delta}{2}\right)\,S_{12}(z_1+z_2)\\
			&\times K_{\ell,2}(z_2)\,S_{12}(z_2-z_1).
		\end{split}
		\label{eq:T2n2}
	\end{equation}
	
	Applying these relations to a complete round trip gives the boundary qKZ equations~\cite{Cherednik1992,ReshetikhinStokmanVlaar2015,PasnooriBoundaryqKZ2026}
	\begin{equation}
		F(\boldsymbol z+\Delta e_j)=\mathcal T_j(\boldsymbol z)F(\boldsymbol z),
		\label{eq:BqKZ}
	\end{equation}
	and their compatibility is the flatness relation
	\begin{equation}
		\mathcal T_i(\boldsymbol z+\Delta e_j)\mathcal T_j(\boldsymbol z) =\mathcal T_j(\boldsymbol z+\Delta e_i)\mathcal T_i(\boldsymbol z).
		\label{eq:BqKZflat}
	\end{equation}
	The order of the shifts matters because transporting particle $j$ changes its label before particle $i$ is transported. The two paths to $\boldsymbol z+\Delta e_i+\Delta e_j$ give the two sides of Eq.~\eqref{eq:BqKZflat}. Yang--Baxter moves reorder the bulk exchanges, braiding unitarity \eqref{eq:Sunitarity} removes a backtracking pair, and the reflection equations reorder the endpoint processes with the appropriate orientation. Hence Eq.~\eqref{eq:BqKZflat} follows from Eqs.~\eqref{eq:YBE} and \eqref{eq:Sunitarity}, with $K_\ell$ satisfying Eq.~\eqref{eq:RE} and $K_\rr$ satisfying Eq.~\eqref{eq:REright}~\cite{Cherednik1992,ReshetikhinStokmanVlaar2015,PasnooriBoundaryqKZ2026}. The two-particle calculation is given in App.~\ref{app:n2}.
	
	\subsection{The autonomous limit}
	\label{sec:autonomouslimit}
	
	As $\kappa\to0$, the labels approach constants. To retain a nontrivial spatial wavefunction, the coefficient function must scale with $\kappa$. In a plane-wave sector,
	\[
	F\sim\exp\left(-i\sum_j k_jz_j/\kappa\right),
	\]
	and the shift $z_j\mapsto z_j+2\kappa\sL$ produces $e^{-2ik_j\sL}$. The boundary qKZ equation then becomes the usual boundary Bethe quantization condition. The $SU(2)$ impurity is treated in Sec.~\ref{sec:smallstep}. The same-chirality arguments tend to zero, so the corresponding auxiliary factors reduce to $S(0)$.
	
	\subsection{Solving the system}
	
	Let $F_{Q_0}^{\mathrm{bqKZ}}(\boldsymbol z)\in\mathcal F_n$ satisfy Eq.~\eqref{eq:BqKZ} in one reference ordering sector. Equation~\eqref{eq:boundarysectortransport} determines the amplitudes in the remaining sectors. Substitution into Eq.~\eqref{eq:boundaryfixedNstate} and the pullback Eq.~\eqref{eq:pullback} then give the spacetime wavefunction.
	
	Off-shell boundary Bethe vectors give explicit solutions of boundary qKZ equations. Bilateral sums and Jackson-type representations are known for diagonal boundaries~\cite{ReshetikhinStokmanVlaar2015,ReshetikhinStokmanVlaar2018}. In the present convention the transport in Eq.~\eqref{eq:doubleRow} has step $\Delta$ and uses $K_\ell(z_j)$ at the left endpoint and $K_\rr(-z_j-\Delta/2)$ at the right endpoint. Since $\widetilde K(u)=K_\rr(-u)$ satisfies Eq.~\eqref{eq:RE}, the two endpoint factors obey the reflection algebra after the orientation is reversed. The transport itself contains the physical reflection matrices.
	
	In the impurity models considered below, the reflection matrices act on the boundary quantum spaces $\cBleft$ and $\cBright$. For the $SU(2)$ impurity of Sec.~\ref{sec:impurity}, the impurity reflection can be absorbed into one additional site with a fixed label, and the known constructions then give explicit coefficient functions $F_{Q_0}^{\mathrm{bqKZ}}(\boldsymbol z)$. We construct them in Sec.~\ref{sec:kondosolution}. Through Eqs.~\eqref{eq:boundarysectortransport}, \eqref{eq:boundaryfixedNstate}, and \eqref{eq:pullback}, they give explicit time-dependent many-body wavefunctions.
	
	The boundary qKZ solution fixes the factorized wavefunction. The inverse matching relations then give the local Hamiltonian with those scattering and reflection data.
	
	\section{Local reconstruction from the matching matrices}
	\label{sec:cayley}
	
	The scattering and reflection matrices determine the jumps of the many-body wavefunction when a scattering surface or an endpoint is crossed. We obtain the local Hamiltonian by solving this matching problem in reverse. The microscopic one- and two-particle matching problem fixes the normalization of the local couplings. App.~\ref{app:transmission} summarizes the Cayley inversion used below.
	
	\subsection{Bulk contact}
	\label{sec:bulkcontact}
	
	For one right mover and one left mover, let $y=x_R-x_L$. Their relative velocity is $2$, so $y<0$ is the incoming region and $y>0$ is the outgoing region. We define the local bulk matrix $V(u)$ through the two-particle matching relation
	\begin{equation}
		S(u)=\bigl(\id+iV(u)\bigr)^{-1}\bigl(\id-iV(u)\bigr).
		\label{eq:bulk-cayley-forward}
	\end{equation}
	We read the bulk interaction directly from the scattering matrix,
	\begin{equation}
		V(u)=i\,[S(u)-\id][S(u)+\id]^{-1},
		\label{eq:bulkCayley}
	\end{equation}
	whenever $-1\notin\Spec S(u)$. For unitary $S$, $V$ is Hermitian. The physical scattering event at time $t$ samples the argument $2u_B(t)$, hence
	\begin{equation}
		\begin{split}
			H_{\rm bulk}(t)={}&2\int_0^{\sL}\dd x\;
			:\psi_{R,a}^\dagger\psi_{L,b}^\dagger\,
			V\bigl(2u_B(t)\bigr)^{ab}{}_{cd}\\
			&\hspace{20mm}\times\psi_{L,d}\psi_{R,c}: .
		\end{split}
		\label{eq:Hbulk}
	\end{equation}
	Here and below, the delta-function interaction is understood in this matching convention. There is no same-chirality contact term because particles with the same chirality never cross dynamically.
	
	\subsection{Cayley inversion at an impurity}
	\label{sec:boundaryinverse}
	
	For a quantum impurity, the one-particle matching problem gives the same Cayley relation. Let $K(u)$ be the impurity scattering matrix and $B(u)$ the corresponding local interaction matrix. Then
	\begin{equation}
		K(u)=\bigl(i\id-B(u)\bigr)^{-1}\bigl(i\id+B(u)\bigr),
		\label{eq:boundary-forward}
	\end{equation}
	and hence
	\begin{equation}
		B(u)=i\,[K(u)-\id][K(u)+\id]^{-1}.
		\label{eq:boundary-inverse}
	\end{equation}
	For unitary $K$, the reconstructed matrix $B$ is Hermitian whenever the inverse in Eq.~\eqref{eq:boundary-inverse} is finite. The left impurity samples $u=u_B(t)$, so
	\begin{equation}
		H_{\partial}(t)=2\,\psi_R^\dagger(0,t)\,B\bigl(u_B(t)\bigr)\,\psi_L(0,t),
		\label{eq:Hboundary}
	\end{equation}
	where the matching relation Eq.~\eqref{eq:boundary-forward} defines the local operator. The derivation and exceptional values of the inverse map are collected in Apps.~\ref{app:transmission} and \ref{app:cayleydomain}.
	
	We need not convert a matrix boundary to a local bilinear. In that case we keep the reflection condition itself,
	\begin{equation}
		\psi_R(0,t)=K_\ell\bigl(u_\ell(t)\bigr)\psi_L(0,t),
		\label{eq:leftfieldBC}
	\end{equation}
	\begin{equation}
		\psi_L(\sL,t)=K_\rr\bigl(u_\rr(t)\bigr)\psi_R(\sL,t),
		\label{eq:rightfieldBC}
	\end{equation}
	with $u_\ell(t)=u_B(t)$ and $u_\rr(t)=-u_B(t)$. For the unitary models considered below, unitarity of $S$ and $K_\alpha$ preserves probability flux across scattering surfaces and endpoints.
	
	\section{Schr\"odinger evolution and the integrability condition}
	\label{sec:schrodinger}
	
	\subsection{The full time-dependent equation}
	
	We obtain the physical wavefunction by pulling the factorized coefficient function back along the affine characteristics. The affine map is essential here. It makes each spectral label constant along its free trajectory, while the difference of the labels at a right--left scattering event depends only on the scattering time. The interaction does not vary arbitrarily with $t$; we evaluate the fixed scattering data at the affine argument $2u_B(t)$. This map also fixes the arguments at the two endpoints.
	
	Within the difference-form chiral setting developed above, let $S$ be a normalized solution of Eqs.~\eqref{eq:YBE} and \eqref{eq:Sunitarity}, obeying the symmetry $S_{21}(u)=S_{12}(u)$, and let $K_\ell$ and $K_\rr$ solve Eqs.~\eqref{eq:RE} and \eqref{eq:REright} together with $K_\alpha(u)K_\alpha(-u)=\id$. For the unitary realizations below, $S$ and $K_\alpha$ are also unitary on the real spectral arguments sampled by the physical processes. The constants $\kappa$, $\nu_R$, and $\nu_L$ fix the characteristic geometry in Eq.~\eqref{eq:affine}.
	
	The inverse matching formulas give the local couplings whenever the corresponding inverses are finite; App.~\ref{app:cayleydomain} collects the exceptional values.
	
	Let $F_Q$ satisfy the bulk exchange relations, the endpoint reflection relations, and the BqKZ equations. Pulling it back along the characteristics gives the physical wavefunction and, away from the exceptional values of App.~\ref{app:cayleydomain}, the local Schr\"odinger equation below.
	
	Inside a configuration-space ordering region, every label satisfies
	\begin{equation}
		\bigl(\partial_t+\chi_j\partial_{x_j}\bigr)\zeta_j(x_j,t)=0,
		\label{eq:label-transport}
	\end{equation}
	because $\partial_t\zeta_j=-\kappa\chi_j$ and $\chi_j\partial_{x_j}\zeta_j=\kappa\chi_j$. Applying the chain rule to Eq.~\eqref{eq:pullback} gives
	\begin{equation}
		i\partial_t\Psi_Q=H_0\Psi_Q, \qquad x_i\ne x_j,\quad x_j\notin\{0,\sL\},
		\label{eq:free-Schrodinger}
	\end{equation}
	and it is the affine labeling that removes all additional derivative terms.
	
	At a right--left scattering event, Eq.~\eqref{eq:bulkEvent} relates the two limiting values of the wavefunction, and the forward Cayley relation in Eq.~\eqref{eq:bulk-cayley-forward} turns this jump into the local bulk interaction with matrix $V$. At an impurity, Eq.~\eqref{eq:boundary-forward} gives the corresponding chirality-changing matching matrix. Since every pair uses one bulk argument, $2u_B(t)$, the full $n$-particle equation is
	\begin{equation}
		i\partial_t\Psi=
		\Bigl[-i\sum_j\chi_j\partial_{x_j}
		+2\sum_{\substack{j<k\\ \chi_j\ne\chi_k}}
		\delta(x_j-x_k)\,V_{jk}\bigl(2u_B(t)\bigr)\Bigr]\Psi ,
		\label{eq:fullNbodySchrodinger}
	\end{equation}
	inside the interval, together with the endpoint matching conditions in Eqs.~\eqref{eq:leftfieldBC}--\eqref{eq:rightfieldBC}. We interpret the contact term in Eq.~\eqref{eq:fullNbodySchrodinger} through the matching relation in Eq.~\eqref{eq:bulk-cayley-forward}. Here we order $V_{jk}$ with the right-moving factor first. There is no term for a same-chirality pair because its relative coordinate is constant, as in Eq.~\eqref{eq:frozenrelative}. For the impurity realizations below, Eq.~\eqref{eq:Hboundary} gives the corresponding local endpoint interaction.

	The local matching relations determine the evolution at scattering events and endpoints. The Yang--Baxter and reflection equations ensure their local consistency, while Eq.~\eqref{eq:BqKZflat} ensures the additional consistency of complete scattering and reflection circuits across the interval. We call the driven problem integrable when these local and global transports are path independent. This condition concerns the full time-dependent evolution and does not require the instantaneous conserved charges of the frozen Hamiltonian to remain conserved.
	
	For a difference-form bulk $S$-matrix and compatible boundary $K$-matrices, the characteristic map fixes the time-dependent spectral arguments. The boundary qKZ equations impose consistency on the resulting finite-interval transport. Pulling a solution back along the characteristics gives the multiparticle wavefunction in each finite-particle sector, and the inverse matching relations determine the local Hamiltonian wherever the Cayley inverse is finite. The bulk and boundary drives share one characteristic spectral coordinate. The following examples give the corresponding interactions and boundary couplings.
	
	\section{Rational \texorpdfstring{$SU(N)$}{SU(N)} data with a matrix boundary}
	\label{sec:sun}
	
	\subsection{Bulk space and scattering matrix}
	
	Let
	\begin{equation}
		\cV=\mathbb C^N, \qquad S_{12}(u)=\frac{iu\,\id_{12}+\PP_{12}}{iu+1}.
		\label{eq:SUN}
	\end{equation}
	This normalization is already unitary for real $u$, since
	\begin{equation}
		S^\dagger S=\frac{(-iu\,\id+\PP_{12})(iu\,\id+\PP_{12})}{1+u^2}=\id ,
		\label{eq:SUNunitary}
	\end{equation}
	using $\PP_{12}^2=\id$. It satisfies $S_{21}=S_{12}$ and $S(u)S(-u)=\id$, and is a function of $\PP_{12}$ alone. On $\Pi_\pm=\tfrac12(\id\pm\PP_{12})$ its eigenvalues are
	\begin{equation}
		s_+(u)=1, \qquad s_-(u)=\frac{iu-1}{iu+1}=\frac{u+i}{u-i},
		\label{eq:SUNeigen}
	\end{equation}
	so the antisymmetric channel has a pole at $u=i$ and a zero at $u=-i$, and $S(0)=\PP_{12}$.
	
	Applying the bulk inversion formula in Eq.~\eqref{eq:bulkCayley} channel by channel, the symmetric channel gives zero while the antisymmetric channel gives $-1/u$. Thus
	\begin{equation}
		V_{12}(u)=-\frac{1}{u}\,\Pi_-=-\frac{1}{2u}\bigl(\id_{12}-\PP_{12}\bigr).
		\label{eq:VSUN}
	\end{equation}
	In the parametrization \eqref{eq:tanparam} this is $a_+=0$ and $a_-=-1/u$, which is used in App.~\ref{app:normalization}.
	
	We take $S^{RL}=S^{LR}=S$ and, by Eq.~\eqref{eq:allexchange}, $\mathcal S^{RR}_{jk}=\mathcal S^{LL}_{jk}=S_{jk}(\zeta_j-\zeta_k)$, which contributes no local term. Substituting Eq.~\eqref{eq:VSUN} into the general bulk Hamiltonian in Eq.~\eqref{eq:Hbulk} at the physical argument $u=2u_B(t)$ gives
	\begin{equation}
		\begin{split}
			H_{\rm bulk}^{SU(N)}(t)=\frac{1}{2u_B(t)}&\int_0^{\sL}\dd x\;
			:\psi_{R,a}^\dagger\psi_{L,b}^\dagger\\
			&\times(\PP_{12}-\id_{12})^{ab}{}_{cd}\,\psi_{L,d}\psi_{R,c}: .
		\end{split}
		\label{eq:HbulkSUN}
	\end{equation}

	\subsection{Two independent matrix boundaries}
	
	For endpoint $\alpha$, take
	\begin{equation}
		K_\alpha(u)=\frac{\xi_\alpha\id_N+iu\,M_\alpha}{\xi_\alpha+iu}\, ,
		\label{eq:Kcnumbernorm}
	\end{equation}
	with $\xi_\alpha\in\mathbb R$, $M_\alpha^\dagger=M_\alpha$, and $M_\alpha^2=\id_N$. At the algebraic level, the parameter $\xi_\alpha$ may be complex. Reality of $\xi_\alpha$ together with Hermiticity of $M_\alpha$ selects the unitary physical representative. The reflection equation requires $M_\alpha^2$ to be proportional to the identity. We use the normalization $M_\alpha^2=\id_N$, with the proportionality absorbed into $\xi_\alpha$. Appendix~\ref{app:matrixRE} gives the short algebraic check. A direct calculation gives
	\begin{equation}
		K_\alpha(u)^\dagger K_\alpha(u)= \frac{(\xi_\alpha\id_N-iuM_\alpha)(\xi_\alpha\id_N+iuM_\alpha)} {(\xi_\alpha-iu)(\xi_\alpha+iu)}=\id_N\, ,
		\label{eq:KcnumberUnitary}
	\end{equation}
	and it also obeys $K_\alpha(u)K_\alpha(-u)=\id_N$, which is the relation $K_\alpha(u)K_\alpha(-u)=\id$. For $M_\alpha=\id_N$ this gives the trivial endpoint $K_\alpha=\id_N$. For the rational bulk matrix \eqref{eq:SUN}, this constant reflection satisfies the reflection equation because $S(u)$ and $S(v)$ are both functions of $\PP_{12}$ and commute. When $\xi_\alpha=0$, the formula reduces to $K_\alpha=M_\alpha$.
	
	This family is closed under $u\to-u$. One has $K_\alpha(-u)=K_\alpha(u)\big|_{\xi_\alpha\to-\xi_\alpha}$, so by App.~\ref{app:matrixRE} it solves Eqs.~\eqref{eq:RE} and \eqref{eq:REright} alike, and it may be used at either endpoint.
	
	For this example, we impose the endpoints directly through their matching conditions. The model is Eq.~\eqref{eq:HbulkSUN} together with
	\begin{equation}
		\psi_{R,a}(0,t)=[K_\ell(u_\ell(t))]_a{}^b\,\psi_{L,b}(0,t),
		\label{eq:SUNmatrixleftBC}
	\end{equation}
	\begin{equation}
		\psi_{L,a}(\sL,t)=[K_\rr(u_\rr(t))]_a{}^b\,\psi_{R,b}(\sL,t),
		\label{eq:SUNmatrixrightBC}
	\end{equation}
	so that
	\begin{equation}
		H_{\rm matrix}(t)=H_0+H_{\rm bulk}^{SU(N)}(t)
		\label{eq:Hmatrix}
	\end{equation}
	with the endpoint conditions imposed by Eqs.~\eqref{eq:SUNmatrixleftBC} and \eqref{eq:SUNmatrixrightBC}. 
	
	The rational $SU(N)$ bulk gives a driven contact interaction proportional to $1/u_B(t)$, while the two matrix boundaries are evaluated at $u_B(t)$ and $-u_B(t)$. The boundary matrices themselves may be chosen independently, but their time dependence is fixed by the affine function that determines the bulk interaction.
	
	\section{Fundamental \texorpdfstring{$SU(N)$}{SU(N)} boundary impurity}
	\label{sec:impurity}
	
	\subsection{Boundary space and impurity reflection matrix}
	
	Let $\cV=\mathbb C^N$ and $\cBleft=\mathbb C^N_0$. We place the impurity at the left endpoint and take
	\begin{equation}
		K_{\ell,a0}(u;d)=S_{a0}(u+d)\,S_{a0}(u-d),
		\label{eq:KSUNimp}
	\end{equation}
	which solves the left reflection equation~\cite{Sklyanin1988,MezincescuNepomechie1992,WangYangCaoShi2015}. Here $d$ is the impurity spectral parameter. We take the right endpoint to be trivial,
	\begin{equation}
		K_\rr(u)=\id,
	\end{equation}
	so that $\psi_L(\sL,t)=\psi_R(\sL,t)$.
	
	Let $\Pi_\pm=(\id\pm \PP_{a0})/2$. Then $K_{\ell,a0}=\Pi_++k(u;d)\Pi_-$ with
	\begin{equation}
		k(u;d)=\frac{\bigl(i(u+d)-1\bigr)\bigl(i(u-d)-1\bigr)} {\bigl(i(u+d)+1\bigr)\bigl(i(u-d)+1\bigr)} =\frac{D(u)+2iu}{D(u)-2iu},
		\label{eq:kSUNimpA}
	\end{equation}
	\begin{equation}
		D(u)=u^2-1-d^2 ,
		\label{eq:kSUNimp}
	\end{equation}
	using $(iu\mp1)^2+d^2=-(D\pm2iu)$. Unitarity of the reflection matrix requires $|k|=1$, which holds whenever $D$ is real, that is whenever $d^2\in\mathbb R$. This permits both real $d$ and purely imaginary $d$. The analogous condition in the $O(3)$ problem is discussed in Sec.~\ref{sec:mixed}.
	
	Applying the impurity inversion formula \eqref{eq:boundary-inverse} to the two channels, the $\Pi_+$ channel gives zero and the $\Pi_-$ channel gives $-2u/D(u)$. Hence
	\begin{equation}
		B_{a0}(u;d)=-\frac{2u}{D(u)}\,\Pi_- =-\frac{u}{u^2-1-d^2}\bigl(\id-\PP_{a0}\bigr).
		\label{eq:GammaSUNimp}
	\end{equation}
	
	\subsection{Resulting Hamiltonian and field condition}
	
	At the left endpoint, $u_\ell(t)=u_B(t)$. Substituting Eq.~\eqref{eq:GammaSUNimp} into the general impurity Hamiltonian \eqref{eq:Hboundary} gives
	\begin{equation}
		H_{\partial,\rm K}(t)=
		\frac{2u_B(t)}{D(u_B(t))}\,
		\psi_R^\dagger(0,t)\bigl(\PP_{a0}-\id\bigr)\psi_L(0,t).
		\label{eq:HKondoboundary}
	\end{equation}
	The driven impurity model is 
	\begin{equation}
		H_{\rm K}(t)=H_0+H_{\rm bulk}^{SU(N)}(t)+H_{\partial,\rm K}(t).
		\label{eq:HKondo}
	\end{equation}
	The bulk and boundary interactions are thus both written naturally in terms of $\PP-\id$. With generators normalized by $\Tr(t^At^B)=\tfrac12\delta^{AB}$, the permutation decomposes as $\PP_{a0}=\tfrac1N\id+2t_a^AT_0^A$. The boundary interaction contains a scalar term and an $SU(N)$ exchange term. The scalar normalization below removes the scalar term and leaves the pure Kondo interaction.
	
	The left reflection matrix gives the field relation
	\begin{equation}
		\psi_{R,a}(0,t)=[K_{\ell,a0}(u_B(t);d)]_a{}^b\,\psi_{L,b}(0,t),
		\label{eq:KondofieldBC}
	\end{equation}
	with impurity operators included. We take the right endpoint to be $K_\rr=\id$.
	
	\subsection{Scalar normalization and pure exchange}
	\label{sec:tracenorm}
	
	An overall scalar phase does not change the reflection equation, but it does change the interaction reconstructed from Eq.~\eqref{eq:boundary-inverse}. We replace
	\begin{equation}
		K_{\ell,a0}(u;d)\longrightarrow K^{\rm exch}_{\ell,a0}(u;d)=\sigma(u)K_{\ell,a0}(u;d)
		\label{eq:KexchSUN}
	\end{equation}
	and choose $\sigma(u)$ so that the identity component of the Cayley inverse vanishes. Applying Eq.~\eqref{eq:boundary-inverse} to $K^{\rm exch}_{\ell,a0}$ then gives
	\begin{equation}
		B^{\rm exch}_{a0}(u;d)=\frac{g_\partial(u)}{2}t_a^AT_0^A,
		\label{eq:BexchSUN}
	\end{equation}
	with
	\begin{equation}
		g_\partial(u)=\frac{2N\left[\sqrt{N^2D(u)^2+4(N^2-1)u^2}-ND(u)\right]}{(N^2-1)u}.
		\label{eq:gpartial}
	\end{equation}
	App.~\ref{app:normalization} derives the scalar phase and the function $g_\partial$. Substituting Eq.~\eqref{eq:BexchSUN} into the general impurity Hamiltonian \eqref{eq:Hboundary}, with $u=u_\ell(t)=u_B(t)$, gives
	\begin{equation}
		H^{\rm exch}_{\partial,\rm K}(t)=g_\partial\bigl(u_\ell(t)\bigr)\bigl(\psi_R^\dagger t^A\psi_L\bigr)(0,t)T_0^A,
		\label{eq:HexchSUN}
	\end{equation}
	a pure Kondo exchange. Boundary scalar factors occur equally on the two sides of the BqKZ flatness relation, so the boundary normalization preserves the transport wherever it is defined.
	
	The fundamental impurity gives an explicit driven $SU(N)$ Kondo model. The bulk interaction is fixed by Eq.~\eqref{eq:HbulkSUN}, while the impurity reflection matrix fixes the exchange coupling $g_\partial(u_B(t))$ through Eq.~\eqref{eq:boundary-inverse}. Once $u_B(t)$ and the fixed impurity parameter $d$ are specified, the time dependence of both interactions is determined.
	
	The preceding equations give the general $SU(N)$ construction. For $SU(2)$ the nested algebra reduces to one family of spin-lowering operators, and the transport can be written with ordinary boundary Bethe vectors. The impurity reflection is scattering from an extra site with fixed label $d$. The extension to $SU(N)$ is discussed in Sec.~\ref{sec:su2extension}.
	
	\section{Explicit boundary qKZ solution for the \texorpdfstring{$SU(2)$}{SU(2)} Kondo impurity}
	\label{sec:kondosolution}
	
	For $N=2$, take the impurity reflection matrix in Eq.~\eqref{eq:KSUNimp} at the left endpoint and $K_\rr=\id$ at the right endpoint. The impurity reflection is rewritten as scattering from a site with fixed label. Off-shell boundary Bethe vectors then give the spin part of the coefficient function, and summing their roots over shifted lattices gives a solution of the finite-difference equations. The roots are not restricted to isolated Bethe values. To control the convergence of the sums below, write the bulk matrix as
	\begin{equation}
		S_{\eta,ij}(u)=\frac{iu\,\id_{ij}+\eta\PP_{ij}}{iu+\eta},
		\label{eq:Seta}
	\end{equation}
	which reduces to Eq.~\eqref{eq:SUN} at $\eta=1$. We take
	\begin{equation}
		K_{\ell,a0}(u;d)=S_{\eta,a0}(u+d)S_{\eta,a0}(u-d).
	\end{equation}
	The Yang--Baxter, reflection, and flatness relations hold algebraically
	for complex $\eta$. The physical sums are defined by first taking
	$\operatorname{Im}\eta>0$ and then taking the boundary value
	$\eta\to1+i0$~\cite{kattel2026integrable}.
	
	\subsection{The impurity as an additional site}
	
	The impurity reflection matrix is a product of two bulk matrices. In the transport \eqref{eq:doubleRow}, the factor $K_{\ell,j}(z_j;d)=S_{j0}(z_j+d)\,S_{j0}(z_j-d)$ stands between the last factor of $\mathcal A_j$ and the first factor of $\mathcal B_j$. Suppose instead that the left reflection were trivial and that one more site, labeled $0$, sat to the left of particle $1$ with the spectral label
	\begin{equation}
		z_0=d .
		\label{eq:impuritysite}
	\end{equation}
	Particle $j$ would meet it exactly there: $\mathcal A_j$ would end with the crossing $S_{j0}(z_j-z_0)$ on the way to the endpoint, and $\mathcal B_j$ would begin with the crossing $S_{j0}(z_j+z_0)$ on the way back. The impurity acts on the electrons as one more site of the transport. It differs from an electron in one respect: it does not move, so it is never carried around the interval, there is no boundary qKZ equation for it, and its label is not shifted. The boundary qKZ system of the Kondo model with $n$ electrons is thus the boundary qKZ system of $n+1$ sites with $K_\ell=K_\rr=\id$, in which only the electron labels $z_1,\ldots,z_n$ are shifted and $z_0=d$ is a fixed parameter. The impurity enters the Bethe ansatz of the static Kondo model in the same way, as an inhomogeneity of the transfer matrix~\cite{AndreiFuruyaLowenstein1983}. With both reflections trivial, every round-trip operator commutes with the total spin of the electrons and the impurity, $\bm S_{\rm tot}=\sum_{j=0}^n\bm s_j$. It suffices to construct highest-weight states. If $M$ spins are flipped relative to the state $|{\Uparrow}\rangle$ in which the impurity and all electrons point up, then $S^z_{\rm tot}=(n+1)/2-M$. For $M=0$ there is no scattering and $F=\mathcal Q(\bm z)|{\Uparrow}\rangle$, where $\mathcal Q$ has period $\Delta$ in every electron label.
	
	These quasi-constant coefficients describe the free motion on the interval. They are functions of $e^{2\pi iz_j/\Delta}$, and for an electron moving to the left at $(x_j,t)$, with $z_j=\kappa(x_j+t)+u_{B,0}$ and $\Delta=2\kappa\sL$,
	\begin{equation}
		e^{2\pi iz_j/\Delta}=e^{2\pi iu_{B,0}/\Delta}\,e^{i\pi(x_j+t)/\sL}.
		\label{eq:freemodes}
	\end{equation}
	Their powers are the free chiral modes of the interval, with wave numbers $\pi m/\sL$.
	
	\subsection{One electron}
	
	For one electron, Eq.~\eqref{eq:BqKZ} reads $F(z+\Delta)=K_\ell(z;d)F(z)$. In the triplet of the electron and the impurity, $K_\ell=\id$ and the amplitude is quasi-constant with period $\Delta$. In the singlet, $K_\ell$ has the eigenvalue
	\begin{equation}
		k(z)=\prod_{\sigma=\pm}\frac{z+\sigma d+i\eta}{z+\sigma d-i\eta},
		\label{eq:ksinglet}
	\end{equation}
	which is Eq.~\eqref{eq:kSUNimpA} at $\eta=1$. The singlet amplitude must satisfy $f(z+\Delta)=k(z)f(z)$. Since $\Gamma(y+1)=y\Gamma(y)$, this first-order difference equation is solved by
	\begin{equation}
		\mathfrak f_K(z)=\prod_{\sigma=\pm}\frac{\Gamma\bigl(1-(z+\sigma d-i\eta)/\Delta\bigr)}{\Gamma\bigl(1-(z+\sigma d+i\eta)/\Delta\bigr)},
		\label{eq:fK}
	\end{equation}
	unique up to a quasi-constant factor. Let $z$ be real and $\eta=1$. For real $d$ the numerator and denominator of each factor are complex conjugates, and for purely imaginary $d$ complex conjugation exchanges the numerator of one factor with the denominator of the other. In both cases $|\mathfrak f_K(z)|=1$, and $\mathfrak f_K$ has neither zeros nor poles on the real axis.
	
	The general solution of the one-electron boundary qKZ equation is 
	\begin{equation}
		F(z)=\sum_{s=-1}^{1}\mathcal Q_{1,s}(z)\,|1,s\rangle+\mathcal Q_{0,0}(z)\,\mathfrak f_K(z)\,|0,0\rangle,
		\label{eq:Fone}
	\end{equation}
	where $|1,s\rangle$ and $|0,0\rangle$ are the triplet and singlet of the electron and the impurity, and the four coefficients $\mathcal Q_{1,s}$ and $\mathcal Q_{0,0}$ are independent functions with period $\Delta$.
	
	The reflection relations fix the two chiral components. A left mover at $(x,t)$ carries $z_+=u_B(t)+\kappa x$. A right mover at $(x,t)$ left the impurity at time $t-x$, with incoming label $z_-=u_B(t-x)=u_B(t)-\kappa x$. Reflection multiplies its amplitude by $K_\ell(z_-;d)$, and the boundary qKZ equation gives $K_\ell(z_-;d)F(z_-)=F(z_-+\Delta)$. Thus
	\begin{equation}
		\begin{gathered}
			\Psi_L(x,t)=F(z_+),\qquad \Psi_R(x,t)=F(z_-+\Delta),\\
			z_\pm=u_B(t)\pm\kappa x .
		\end{gathered}
		\label{eq:oneelectron}
	\end{equation}
	At $x=0$, $z_+=z_-=u_B(t)$, and Eq.~\eqref{eq:BqKZ} gives $\Psi_R(0,t)=K_\ell(u_B(t);d)\Psi_L(0,t)$. At $x=\sL$, $z_-+\Delta=z_+$, so $\Psi_R(\sL,t)=\Psi_L(\sL,t)$, as required by $K_\rr=\id$. Inside the interval $\Psi_L$ depends only on $x+t$ and $\Psi_R$ only on $x-t$, so the free equations of motion hold.
	
	Equation~\eqref{eq:oneelectron} gives the internal electron--impurity amplitudes. Let $|x,L\rangle$ and $|x,R\rangle$ denote position--chirality states, with the internal spin spaces suppressed. The complete one-electron state is
	\begin{widetext}
		\begin{equation}
			\begin{split}
				|\Psi(t)\rangle={}&\mathcal N\int_0^{\sL}\!\dd x\;\Bigg\{
				|x,L\rangle\otimes\left[
				\sum_{s=-1}^{1}\mathcal Q_{1,s}(z_+)|1,s\rangle
				+\mathcal Q_{0,0}(z_+)\mathfrak f_K(z_+)|0,0\rangle\right]\\
				&\hspace{25mm}+|x,R\rangle\otimes\left[
				\sum_{s=-1}^{1}\mathcal Q_{1,s}(z_-)|1,s\rangle
				+\mathcal Q_{0,0}(z_-)\mathfrak f_K(z_-+\Delta)|0,0\rangle\right]\Bigg\}.
			\end{split}
			\label{eq:fulloneelectron}
		\end{equation}
	\end{widetext}
	As $x$ runs over $[0,\sL]$, the arguments $z_+$ and $z_-+\Delta$ in Eq.~\eqref{eq:oneelectron} together cover the window $[u_B(t),u_B(t)+\Delta]$, which is exactly one period. The state at time $t$ is $F$ on this window, with the left movers on its first half and the right movers on its second half. At $\eta=1$, $|\mathfrak f_K|=1$ on the real axis, so $|F(z)|^2$ has period $\Delta$, and
	\begin{equation}
		\mathcal N^{-2}=\frac{1}{\kappa}\int_{0}^{\Delta}\!\dd z\,\Bigl[\sum_{s=-1}^{1}|\mathcal Q_{1,s}(z)|^2+|\mathcal Q_{0,0}(z)|^2\Bigr]
		\label{eq:onenorm}
	\end{equation}
	does not depend on $t$, as unitarity requires.
	
	The quasi-constants are fixed by the initial state. At $t=0$ the relevant window is $[u_{B,0},u_{B,0}+\Delta]$. The triplet components of the initial data determine $\mathcal Q_{1,s}$, while the singlet component divided by the nonvanishing factor $\mathfrak f_K$ determines $\mathcal Q_{0,0}$. Periodic extension gives the solution at later times. Conversely, any choice of the four coefficient functions with period $\Delta$ gives a solution. A single Fourier sector has
	\[
	\mathcal Q_{1,s}(z)=A_{1,s}e^{2\pi imz/\Delta},
	\qquad
	\mathcal Q_{0,0}(z)=A_{0,0}e^{2\pi imz/\Delta},
	\]
	with $m\in\mathbb Z$. Since $e^{2\pi iz_\pm/\Delta}=e^{2\pi iu_{B,0}/\Delta}e^{i\pi(t\pm x)/\sL}$, the triplet components have the free interval modes with wave numbers $\pi m/\sL$. Direct propagation along the characteristics gives this result for arbitrary initial data, with $K_\ell(u_B(t);d)$ applied at each arrival at $x=0$ and $K_\rr=\id$ at $x=\sL$.
	
	The Gamma-function ratio collects the reflection history of the electron. Between two reflections at the impurity, the electron travels once around the interval, a time $2\sL$, and its label advances by $\Delta$. Iterating the difference equation $m$ times and using Stirling's formula for the remaining factor gives
	\begin{equation}
		\mathfrak f_K(z)=\lim_{m\to\infty}m^{4i\eta/\Delta}\prod_{k=1}^{m}k(z-k\Delta).
		\label{eq:Kprod}
	\end{equation}
	Each factor is the singlet reflection phase of an earlier reflection at the impurity. The boundary coupling of Eq.~\eqref{eq:GammaSUNimp} decays as $1/u_B(t)$, so these phases decrease as $1/k$ and add up to a logarithm, which the factor $m^{4i\eta/\Delta}$ removes. This logarithm also appears at late times. The reflection formula $\Gamma(y)\Gamma(1-y)=\pi/\sin(\pi y)$ and Stirling's formula give
	\begin{equation}
		\mathfrak f_K(z)=\mathfrak s(z)\Bigl(\frac{z}{\Delta}\Bigr)^{4i\eta/\Delta}\bigl[1+O(\Delta/z)\bigr],
		\label{eq:Klate}
	\end{equation}
	with the quasi-constant factor
	\begin{equation}
		\mathfrak s(z)=\prod_{\sigma=\pm}\frac{\sin\bigl[\pi(z+\sigma d+i\eta)/\Delta\bigr]}{\sin\bigl[\pi(z+\sigma d-i\eta)/\Delta\bigr]},
	\end{equation}
	of unit modulus for real $z$. Since $z/\Delta\simeq t/2\sL$ at late times, the singlet of the electron and the impurity acquires the phase $(t/2\sL)^{2i/(\kappa\sL)}$ at $\eta=1$, a Coulomb-type logarithm set by the drive rate and the length of the interval.
	
	\subsection{Boundary off-shell Bethe vectors}
	
	The impurity reflection is incorporated by treating the impurity as site $0$ with fixed label $z_0=d$. The physical round-trip operator in Eq.~\eqref{eq:doubleRow} still contains the reflection matrices of the electron. For $K_\rr=\id$, the round trip of the last electron can also be written as the trace of an auxiliary double-row monodromy.
	
	Let $a$ be an auxiliary spin-$\tfrac12$ and define
	\begin{equation}
		\begin{split}
			\mathbb T_a(v)&=S_{a0}(v-z_0)\,S_{a1}(v-z_1)\cdots S_{an}(v-z_n),\\
			\mathbb U_a(\theta)&=\mathbb T_a(\theta)^{-1}\,\mathbb T_a(-\theta),
		\end{split}
		\label{eq:doublerow}
	\end{equation}
	with each bulk factor $S_{ij}=S_{\eta,ij}$ and $z_0=d$. By braiding unitarity, $\mathbb T_a(\theta)^{-1}=S_{an}(z_n-\theta)\cdots S_{a0}(z_0-\theta)$, so $\mathbb U_a$ is a double row: the auxiliary spin passes the sites $n,\ldots,0$ in $\mathbb T_a(-\theta)$ and returns through $0,\ldots,n$. In the auxiliary basis $(|\uparrow\rangle_a,|\downarrow\rangle_a)$ we write
	\begin{equation}
		\mathbb U_a(\theta)=\begin{pmatrix}\mathsf A(\theta)&\mathsf B(\theta)\\ \mathsf C(\theta)&\mathsf D(\theta)\end{pmatrix}_a .
	\end{equation}
	The trace acts on an off-shell Bethe vector in two parts. One preserves
	the set of roots and is proportional to the original vector; the other
	replaces one root by the spectral parameter of the trace. The ordinary
	Bethe ansatz removes the second term by imposing equations on the roots.
	Here the roots are summed over a lattice, and the same term cancels
	between neighboring lattice points. The calculation is given in
	App.~\ref{app:bqkzproof}.
	
	On the state $|{\Uparrow}\rangle$ in which the impurity and all electrons point up,
	\begin{equation}
		\mathsf A(\theta)|{\Uparrow}\rangle=|{\Uparrow}\rangle,\qquad
		\mathsf C(\theta)|{\Uparrow}\rangle=0,\qquad
		\mathsf D(\theta)|{\Uparrow}\rangle=\mathfrak d(\theta)|{\Uparrow}\rangle,
		\label{eq:bdyvacuum}
	\end{equation}
	with
	\begin{equation}
		\begin{gathered}
			\mathfrak d(\theta)=\frac{2\theta}{2\theta+i\eta}\,\mathfrak g(\theta)+\frac{i\eta}{2\theta+i\eta},\\
			\mathfrak g(\theta)=\prod_{r=0}^{n}\frac{(\theta-z_r)(\theta+z_r)}{(\theta-z_r+i\eta)(\theta+z_r+i\eta)} .
		\end{gathered}
		\label{eq:bdyvacuumeig}
	\end{equation}
	Besides the product $\mathfrak g$ of diagonal amplitudes, $\mathsf D$ receives contributions from paths in which the auxiliary spin flips one site on the way out and flips it back on the way in. Their sum telescopes to the form above; App.~\ref{app:bqkzproof} gives the derivation. The entry $\mathsf B(\theta)$ lowers $S^z_{\rm tot}$ by one. We use the renormalized operator (compare Ref.~\cite{ReshetikhinStokmanVlaar2015})
	\begin{equation}
		\begin{gathered}
			\bar{\mathsf B}(\theta)=\Upsilon(\theta)\,\frac{2\theta}{2\theta-i\eta}\;\mathsf B\bigl(\theta-\tfrac{i\eta}{2}\bigr),\\
			\Upsilon(\theta)=\prod_{r=0}^n\frac{\theta-z_r+i\eta/2}{\theta-z_r-i\eta/2},
		\end{gathered}
		\label{eq:Bbar}
	\end{equation}
	and the off-shell boundary Bethe vectors
	\begin{equation}
		\Omega(\bm z;\bm\theta)=\bar{\mathsf B}(\theta_1)\cdots\bar{\mathsf B}(\theta_M)\,|{\Uparrow}\rangle ,
		\label{eq:bdyoffshell}
	\end{equation}
	with unrestricted roots $\theta_1,\ldots,\theta_M$. The operators $\bar{\mathsf B}(\theta)$ commute with each other. For one root the vector takes an explicit form. Writing $\varphi_r(v)\sigma_r^-|{\Uparrow}\rangle$ for the components of the ordinary single-row vector $\langle\uparrow|_a\mathbb T_a(v)|\downarrow\rangle_a|{\Uparrow}\rangle$,
	\begin{equation}
		\varphi_r(v)=\frac{-i\eta}{v-z_r-i\eta}\prod_{l>r}\frac{v-z_l}{v-z_l-i\eta},\qquad r=0,\ldots,n,
		\label{eq:phibdy}
	\end{equation}
	one finds
	\begin{equation}
		\begin{gathered}
			\bar{\mathsf B}(\theta)|{\Uparrow}\rangle=\mathcal E(\theta)-\mathcal E(-\theta),\\
			\mathcal E(\theta)=\Upsilon(\theta)\sum_{r=0}^n\varphi_r\bigl(-\theta+\tfrac{i\eta}{2}\bigr)\sigma_r^-|{\Uparrow}\rangle .
		\end{gathered}
		\label{eq:Bbarexplicit}
	\end{equation}
	Since $\varphi_r(-\theta+i\eta/2)=[i\eta/(\theta+z_r+i\eta/2)]\prod_{l>r}(\theta+z_l-i\eta/2)/(\theta+z_l+i\eta/2)$, the component of $\mathcal E(\theta)$ along $\sigma_r^-|{\Uparrow}\rangle$ is
	\begin{equation}
		e_r(\theta)=\frac{i\eta\,\Upsilon(\theta)}{\theta+z_r+i\eta/2}\prod_{l>r}\frac{\theta+z_l-i\eta/2}{\theta+z_l+i\eta/2}.
		\label{eq:ek}
	\end{equation}
	The vector is odd under $\theta\to-\theta$. This sign change records the reflection of the root at the endpoint.
	
	The trace of the double row, $\mathsf t(v)=\Tr_a\mathbb U_a(v)=\mathsf A(v)+\mathsf D(v)$, acts on a one-root vector as
	\begin{equation}
		\begin{split}
			\mathsf t(v)\bar{\mathsf B}(\theta)|{\Uparrow}\rangle={}&\mathscr L(v;\theta)\,\bar{\mathsf B}(\theta)|{\Uparrow}\rangle\\
			&+\frac{2i\eta(v+i\eta)\,[\Upsilon(\theta)-\Upsilon(-\theta)]}{(v-\theta+i\eta/2)(v+\theta+i\eta/2)}\,\mathsf B(v)|{\Uparrow}\rangle ,
		\end{split}
		\label{eq:bdyoffshellaction}
	\end{equation}
	where $\mathscr L(v;\theta)$ is given in Eq.~\eqref{eq:wantedL}. The first term is the wanted term: it contains the original vector. The second term is unwanted: the root has been replaced by the spectral parameter $v$. Its coefficient contains $\Upsilon(\theta)-\Upsilon(-\theta)$, which plays the role of $a(\zeta)-d(\zeta)$ in the periodic problem~\cite{kattel2026integrable}. The ordinary Bethe ansatz sets this combination to zero,
	\begin{equation}
		\prod_{r=0}^{n}\prod_{\upsilon=\pm}\frac{\theta+\upsilon z_r+i\eta/2}{\theta+\upsilon z_r-i\eta/2}=1,
	\end{equation}
	which is the one-root Bethe equation of the open chain with site labels $z_0,\ldots,z_n$. For $M$ roots, the unwanted term in which $\theta_m$ is replaced by $v$ has a coefficient proportional to $\mathcal X_m(\theta_m)-\mathcal X_m(-\theta_m)$, where
	\begin{equation}
		\begin{gathered}
			\mathcal X_m(\theta_m)=\Upsilon(\theta_m)\prod_{k\ne m}\mathfrak r(\theta_m,\theta_k),\\
			\mathfrak r(a,b)=\frac{(a-b-i\eta)(a+b-i\eta)}{(a-b)(a+b)},
		\end{gathered}
		\label{eq:Xm}
	\end{equation}
	and only $\theta_m$ changes sign in $\mathcal X_m(-\theta_m)$.
	
	The transfer matrix enters the transport through the last electron. Its return leg $\mathcal C_n$ is empty and $K_\rr=\id$, so $\mathcal T_n$ contains no $\Delta$. Evaluating the double row at $\theta=-z_n$, where $S_{an}(0)=\PP_{an}$, gives (App.~\ref{app:bqkzproof})
	\begin{equation}
		\mathcal T_n(\bm z)=\frac{2z_n-i\eta}{2(z_n-i\eta)}\,\mathsf t(-z_n).
		\label{eq:Tntransfer}
	\end{equation}
	This is the boundary counterpart of $\mathcal K_1=\mathsf t(\xi_1)/a(\xi_1)$ in the periodic problem~\cite{kattel2026integrable}. It is an operator identity that uses only $S(0)=\PP$ and braiding unitarity, and it holds for every $\Delta$. The trace in $\mathsf t$ is closed with the identity, which is the dual reflection matrix belonging to $K_\rr=\id$; the transport contains $K_\rr=\id$ at the same place. Two conditions make the identity possible. The return leg of the last electron is empty, and the right reflection is trivial. For $j<n$ the return leg contains the shifted label $z_j+\Delta$, and for a nontrivial $K_\rr$ the transport contains $K_\rr(-z_n-\Delta/2)$. In either case the round trip is in general not Sklyanin's transfer matrix; for a nontrivial, nonconstant $K_\rr$ the matrix needed to close the trace depends on $\Delta$ and does not solve the dual reflection equation (App.~\ref{app:bqkzproof}). The round trips $j<n$ are obtained from the exchange relation instead, and a nontrivial far reflection has to be carried by the weight, as in Ref.~\cite{ReshetikhinStokmanVlaar2015}. At $v=-z_n$ the second term of $\mathfrak d$ drops out, the prefactor in Eq.~\eqref{eq:Tntransfer} cancels the coefficient of $\mathsf A$, and $\mathsf B(-z_n)|{\Uparrow}\rangle=[2z_n/(2z_n-i\eta)]\sigma_n^-|{\Uparrow}\rangle$ flips only the spin of the transported electron. Hence
	\begin{equation}
		\mathcal T_n(\bm z)\bar{\mathsf B}(\theta)|{\Uparrow}\rangle=\mathscr L_n(\theta)\,\bar{\mathsf B}(\theta)|{\Uparrow}\rangle+\mathscr U_n(\theta)\,\sigma_n^-|{\Uparrow}\rangle,
		\label{eq:Tnaction}
	\end{equation}
	with
	\begin{equation}
		\begin{gathered}
			\mathscr L_n(\theta)=\frac{(\theta+z_n+i\eta/2)(\theta-z_n-i\eta/2)}{(\theta+z_n-i\eta/2)(\theta-z_n+i\eta/2)},\\
			\mathscr U_n(\theta)=\frac{2i\eta z_n\,[\Upsilon(\theta)-\Upsilon(-\theta)]}{(\theta+z_n-i\eta/2)(\theta-z_n+i\eta/2)} .
		\end{gathered}
		\label{eq:LnUn}
	\end{equation}
	
	\subsection{Summing over the roots}
	
	For one root, take the lattice $\Lambda=\theta^0+\Delta\mathbb Z$ with a generic origin $\theta^0$ and form the coefficient function as a weighted sum of off-shell vectors. The scalar weight is fixed by the transport equation:
	\begin{equation}
		F(\bm z)=\sum_{\theta\in\Lambda}\Phi_1(\bm z;\theta)\,\bar{\mathsf B}(\theta)|{\Uparrow}\rangle .
		\label{eq:bdyFone}
	\end{equation}
	We require $F(\bm z+\Delta e_n)=\mathcal T_n(\bm z)F(\bm z)$. The left side changes both the weight and the vector. In every component $\sigma_r^-|{\Uparrow}\rangle$ with $r<n$, the label $z_n$ enters $e_r(\pm\theta)$ only through two factors, and the shift of these factors combined with the shift of the weight found below gives exactly $\mathscr L_n(\theta)$. These components agree with Eq.~\eqref{eq:Tnaction} term by term (App.~\ref{app:bqkzproof}). The unwanted term has only the component $r=n$. In that component, the difference $F(\bm z+\Delta e_n)-\mathcal T_n(\bm z)F(\bm z)$ equals $i\eta(2z_n+\Delta)$ times
	\begin{equation}
		\begin{split}
			&\sum_{\theta\in\Lambda}\Phi_1(\bm z;\theta)\frac{\Upsilon(-\theta)}{(\theta+z_n-i\eta/2)(\theta-z_n-\Delta+i\eta/2)}\\
			&\quad-\sum_{\theta\in\Lambda}\Phi_1(\bm z;\theta)\frac{\Upsilon(\theta)}{(\theta-z_n+i\eta/2)(\theta+z_n+\Delta-i\eta/2)}.
		\end{split}
		\label{eq:bdyunwanted}
	\end{equation}
	The second sum contains $\Upsilon(\theta)$ at one lattice point. To compare it with the first sum, shift that point to its neighbor $\theta+\Delta$. The denominator becomes
	\begin{equation}
		\bigl[(\theta+\Delta)-z_n-\Delta+i\eta/2\bigr]\bigl[(\theta+\Delta)+z_n-i\eta/2\bigr],
	\end{equation}
	which is the denominator of the first sum evaluated at $\theta+\Delta$. The numerators match after this shift if the one-root weight obeys
	\begin{equation}
		\Phi_1(\bm z;\theta)\,\Upsilon(\theta)=\Phi_1(\bm z;\theta+\Delta)\,\Upsilon(-\theta-\Delta).
		\label{eq:bdykeyid}
	\end{equation}
	Inserting Eq.~\eqref{eq:bdykeyid} into the second sum gives
	\begin{equation}
		\begin{split}
			\sum_{\theta\in\Lambda}&\Phi_1(\bm z;\theta+\Delta)\,\Upsilon\bigl(-(\theta+\Delta)\bigr)\\
			&\times\frac{1}{\bigl[(\theta+\Delta)+z_n-\frac{i\eta}{2}\bigr]\bigl[(\theta+\Delta)-z_n-\Delta+\frac{i\eta}{2}\bigr]}.
		\end{split}
	\end{equation}
	The change of variable $\theta'=\theta+\Delta$ leaves $\Lambda$ unchanged, so this is the first sum in Eq.~\eqref{eq:bdyunwanted}. The unwanted contribution cancels. Compared with the periodic identity $\Phi a(\zeta)=\Phi(\zeta+\varpi)d(\zeta+\varpi)$, the factor $a$ is replaced by $\Upsilon(\theta)$ and the factor $d$ by its mirror image $\Upsilon(-\theta')$ at the neighboring point: the shift moves a root to its neighbor, and the reflection replaces it by its mirror image. The lattice need not be symmetric under $\theta\to-\theta$.
	
	Dividing Eq.~\eqref{eq:bdykeyid} by $\Phi_1(\bm z;\theta)\Upsilon(-\theta-\Delta)$ gives the first-order difference equation
	\begin{equation}
		\frac{\Phi_1(\bm z;\theta+\Delta)}{\Phi_1(\bm z;\theta)}=\prod_{r=0}^{n}\frac{\theta-z_r+i\eta/2}{\theta-z_r-i\eta/2}\,\frac{\theta+z_r+\Delta+i\eta/2}{\theta+z_r+\Delta-i\eta/2}.
		\label{eq:bdyPhirec}
	\end{equation}
	Each site contributes two factors, one for the root and one for its mirror image. We seek a function $\mathcal W(x)$ with
	\begin{equation}
		\frac{\mathcal W(x+\Delta)}{\mathcal W(x)}=\frac{x+\Delta+i\eta/2}{x+\Delta-i\eta/2}.
		\label{eq:bdyWshift}
	\end{equation}
	Then $\mathcal W(z_r+\theta)$ supplies the second factor of Eq.~\eqref{eq:bdyPhirec}, and, since Eq.~\eqref{eq:bdyWshift} also gives $\mathcal W(x-\Delta)/\mathcal W(x)=(x-i\eta/2)/(x+i\eta/2)$, the factor $\mathcal W(z_r-\theta)$ supplies the first. With $y=x/\Delta$ and $\Gamma(y+1)=y\Gamma(y)$, a solution is
	\begin{equation}
		\mathcal W(x)=\frac{\Gamma\bigl(x/\Delta+1+i\eta/2\Delta\bigr)}{\Gamma\bigl(x/\Delta+1-i\eta/2\Delta\bigr)},
		\label{eq:bdyW}
	\end{equation}
	so that
	\begin{equation}
		\Phi_1(\bm z;\theta)=\prod_{r=0}^{n}\mathcal W(z_r+\theta)\,\mathcal W(z_r-\theta).
		\label{eq:bdyPhione}
	\end{equation}
	A factor with period $\Delta$ in $\theta$ leaves Eq.~\eqref{eq:bdyPhirec} unchanged and belongs to the quasi-constant freedom. The same weight makes the components $r<n$ agree: under $z_n\to z_n+\Delta$ the two factors $\mathcal W(z_n\pm\theta)$ change by $(\theta+z_n+\Delta+i\eta/2)/(\theta+z_n+\Delta-i\eta/2)$ and $(\theta-z_n-\Delta-i\eta/2)/(\theta-z_n-\Delta+i\eta/2)$, and these cancel every $\Delta$-dependent factor produced by the shift of $e_r(\pm\theta)$.
	
	For several roots there is a second ingredient. The unwanted coefficients contain the functions $\mathfrak r(\theta_m,\theta_k)$ of Eq.~\eqref{eq:Xm}, and these must also be transported when one root is shifted. We introduce the pair weight
	\begin{equation}
		\mathcal G(x)=\frac{x}{\Delta}\,\frac{\Gamma\bigl(x/\Delta-i\eta/\Delta\bigr)}{\Gamma\bigl(x/\Delta+1+i\eta/\Delta\bigr)},
		\label{eq:bdyG}
	\end{equation}
	which obeys
	\begin{equation}
		\frac{\mathcal G(x+\Delta)}{\mathcal G(x)}=\frac{x+\Delta}{x}\,\frac{x-i\eta}{x+\Delta+i\eta},
		\label{eq:bdyGshift}
	\end{equation}
	and the complete weight
	\begin{equation}
		\begin{split}
			\Phi(\bm z,\bm\theta)={}&\prod_{r=0}^n\prod_{m=1}^M\mathcal W(z_r+\theta_m)\mathcal W(z_r-\theta_m)\\
			&\times\prod_{m<m'}\mathcal G(\theta_m+\theta_{m'})\mathcal G(\theta_m-\theta_{m'}).
		\end{split}
		\label{eq:bdyPhi}
	\end{equation}
	Each root enters the weight through $\mathcal W(z_r\pm\theta_m)$, that is, through its separation from every label and from the mirror image of every label at the left endpoint, and each pair of roots through $\mathcal G$ at their difference and at their sum. Applying Eq.~\eqref{eq:bdyGshift} to $\mathcal G(\theta_m\pm\theta_k)$ gives, for both orderings of $m$ and $k$, the ratio $\mathfrak r(\theta_m,\theta_k)/\mathfrak r(-\theta_m-\Delta,\theta_k)$. Together with the $\mathcal W$ factors this gives the multi-root form of Eq.~\eqref{eq:bdykeyid},
	\begin{equation}
		\Phi(\bm z,\bm\theta)\,\mathcal X_m(\theta_m)=\Phi(\bm z,\bm\theta+\Delta e_m)\,\mathcal X_m(-\theta_m-\Delta),
		\label{eq:bdykeyidM}
	\end{equation}
	which converts the unwanted term of root $m$ into a difference between neighboring lattice points in the $\theta_m$ direction.
	
	For a lattice $\Lambda=\bm\theta^0+\Delta\mathbb Z^M$ with generic $\bm\theta^0$, we define
	\begin{equation}
		F_\Lambda(\bm z)=\sum_{\bm\theta\in\Lambda}\Phi(\bm z,\bm\theta)\,\Omega(\bm z;\bm\theta),
		\label{eq:bdyjackson}
	\end{equation}
	with $z_0=d$ in both $\Phi$ and $\Omega$. These sums have the structure of the bilateral sums of Ref.~\cite{ReshetikhinStokmanVlaar2015} for diagonal reflection matrices, here with both reflections trivial after the impurity is absorbed into the site $0$. The proof below and in App.~\ref{app:bqkzproof} does not rely on that correspondence. For one root, the argument above proves $F_\Lambda(\bm z+\Delta e_n)=\mathcal T_n(\bm z)F_\Lambda(\bm z)$. When the labels of two neighboring electrons are interchanged, $F_\Lambda$ obeys
	\begin{equation}
		F_\Lambda(\ldots,z_{a+1},z_a,\ldots)=\PP_{a,a+1}S_{a,a+1}(z_{a+1}-z_a)\,F_\Lambda(\bm z),
		\label{eq:bdybosonic}
	\end{equation}
	with the same argument as the same-chirality factors of $\mathcal A_j$. This exchange relation carries the equation for $j=n$ to every $j$ (App.~\ref{app:bqkzproof}). Thus $F_\Lambda$ solves Eq.~\eqref{eq:BqKZ} for every electron $j=1,\ldots,n$. It is a highest-weight vector, $S^+_{\rm tot}F_\Lambda=0$ and $S^z_{\rm tot}F_\Lambda=[(n+1)/2-M]F_\Lambda$.
	
	For one flipped spin, Eqs.~\eqref{eq:ek} and~\eqref{eq:bdyPhione} give the fully explicit amplitude. Its component with the spin of site $r$ flipped is
	\begin{equation}
		\begin{split}
			F_{\Lambda,r}(\bm z)={}&i\eta\sum_{\theta\in\Lambda}\Phi_1(\bm z;\theta)\biggl[\frac{\Upsilon(\theta)}{\theta+z_r+i\eta/2}\\
			&\times\prod_{l>r}\frac{\theta+z_l-i\eta/2}{\theta+z_l+i\eta/2}-(\theta\to-\theta)\biggr],
		\end{split}
		\label{eq:bdyonemagnon}
	\end{equation}
	for $r=0,\ldots,n$. For one electron this sum is the singlet $|0,0\rangle$ times $\mathfrak f_K(z)$ of Eq.~\eqref{eq:fK}, up to a quasi-constant factor.
	
	Fermionic statistics fixes the remaining freedom. In the reference configuration all electrons are left movers, and exchanging two of them requires $F(\ldots,z_{a+1},z_a,\ldots)=-\PP_{a,a+1}S_{a,a+1}(z_{a+1}-z_a)F(\bm z)$. Since Eq.~\eqref{eq:bdybosonic} has the opposite sign, the sums are multiplied by quasi-constant coefficients that are antisymmetric in the electron labels, for example $\prod_{1\le a<b\le n}(e^{2\pi iz_a/\Delta}-e^{2\pi iz_b/\Delta})$. Together with the states obtained by lowering, this gives solutions in every spin sector. Whether these sums span all solutions for the present $SU(2)$-invariant boundaries is not established here; Ref.~\cite{ReshetikhinStokmanVlaar2018} proves the corresponding statement for diagonal boundaries.
	
	\subsection{Boundary of convergence and the physical coupling}
	\label{sec:bdyconvergence}
	
	We first discuss the convergence prescription for the one-root contribution in Eq.~\eqref{eq:bdyonemagnon}. The spectral lattice is
	\begin{equation}
		\theta_k=\theta^0+k\Delta,\qquad k\in\mathbb Z.
		\label{eq:bdylattice}
	\end{equation}
	We keep the scalar weight and the off-shell vector together and define
	\begin{equation}
		Y_k=\Phi_1(\bm z;\theta_k)\,\bar{\mathsf B}(\theta_k)|\Uparrow\rangle.
		\label{eq:bdyYk}
	\end{equation}
	Thus $Y_k$ is one vector-valued summand of the lattice sum. Whenever the sum converges, the one-root amplitude is
	\begin{equation}
		F_\Lambda(\bm z)=\sum_{k\in\mathbb Z}Y_k.
		\label{eq:bdyfullsum}
	\end{equation}
	
	The scalar part of the summand is
	\begin{equation}
		\Phi_1(\bm z;\theta)=\prod_{r=0}^{n}\mathcal W(z_r+\theta)\mathcal W(z_r-\theta),
		\label{eq:bdyPhiOne}
	\end{equation}
	where
	\begin{equation}
		\mathcal W(x)=\frac{\Gamma\left(1+\frac{x}{\Delta}+\frac{i\eta}{2\Delta}\right)}{\Gamma\left(1+\frac{x}{\Delta}-\frac{i\eta}{2\Delta}\right)}.
		\label{eq:bdyW-final}
	\end{equation}
	
	For large positive $X$,
	\begin{equation}
		\frac{\Gamma(X+a)}{\Gamma(X+b)}=X^{a-b}\left[1+O(X^{-1})\right].
		\label{eq:gammaratio}
	\end{equation}
	It follows that, for $k\to+\infty$,
	\begin{equation}
		\mathcal W(z_r\pm\theta_k)=C_{r,\pm}^{(+)}k^{i\eta/\Delta}\left[1+O(k^{-1})\right],
		\label{eq:bdyWplus}
	\end{equation}
	where $C_{r,\pm}^{(+)}$ is independent of $k$.
	
	For $k\to-\infty$, the Gamma-function arguments become large and negative. Using
	\begin{equation}
		\Gamma(X)\Gamma(1-X)=\frac{\pi}{\sin(\pi X)},
		\label{eq:gammareflection}
	\end{equation}
	we rewrite them in terms of Gamma functions with large positive arguments. The sine factors are bounded on the lattice, and the factors of $(-1)^k$ cancel between numerator and denominator. Thus,
	\begin{equation}
		\mathcal W(z_r\pm\theta_k)=C_{r,\pm}^{(-)}|k|^{i\eta/\Delta}\left[1+O(|k|^{-1})\right],\qquad k\to-\infty.
		\label{eq:bdyWminus}
	\end{equation}
	
	There are two such factors for each of the $n+1$ sites. Hence
	\begin{equation}
		\Phi_1(\bm z;\theta_k)=C_\Phi^{(\pm)}|k|^{2i(n+1)\eta/\Delta}\left[1+O(|k|^{-1})\right],\qquad k\to\pm\infty.
		\label{eq:bdyPhiasymptotic}
	\end{equation}
	
	For the off-shell vector, the large-$|\theta|$ behavior follows from Eq.~\eqref{eq:ek},
	\begin{equation}
		e_r(\theta)=\frac{i\eta}{\theta}+O(\theta^{-2}).
		\label{eq:bdyerasy}
	\end{equation}
	Therefore,
	\begin{equation}
		\bar{\mathsf B}(\theta)|\Uparrow\rangle=\frac{2i\eta}{\theta}\sum_{r=0}^{n}\sigma_r^-|\Uparrow\rangle+O(\theta^{-2}).
		\label{eq:bdyBasy}
	\end{equation}
	Since $\theta_k=\theta^0+k\Delta$, this vector is of order $|k|^{-1}$. Thus there are constant vectors $C_+$ and $C_-$ such that
	\begin{equation}
		Y_k=C_+k^{\rho-1}\left[1+O(k^{-1})\right],\qquad k\to+\infty,
		\label{eq:bdyYplus}
	\end{equation}
	and
	\begin{equation}
		Y_{-k}=C_-k^{\rho-1}\left[1+O(k^{-1})\right],\qquad k\to+\infty,
		\label{eq:bdyYminus}
	\end{equation}
	where
	\begin{equation}
		\rho=\frac{2i(n+1)\eta}{\Delta}.
		\label{eq:bdyrho}
	\end{equation}
	
	At the physical value $\eta=1$, one has $\operatorname{Re}\rho=0$. The terms then decrease as $1/|k|$ while their phase continues to rotate. The lattice sum lies on the boundary of convergence. We define it first for $\operatorname{Im}\eta>0$, where
	\begin{equation}
		\operatorname{Re}\rho=-\frac{2(n+1)\operatorname{Im}\eta}{\Delta}<0.
		\label{eq:bdyReRho}
	\end{equation}
	Both tails then converge absolutely. The physical value is obtained by continuing to
	\begin{equation}
		\eta\longrightarrow1+i0.
		\label{eq:bdyphysicaleta}
	\end{equation}
	
	To make the cutoff dependence explicit, define the symmetric partial sum
	\begin{equation}
		S_K=\sum_{|k|\le K}Y_k=\sum_{k=-K}^{K}Y_k.
		\label{eq:bdypartialdef}
	\end{equation}
	The full sum is decomposed as
	\begin{equation}
		F_\Lambda=\sum_{|k|\le K}Y_k+\sum_{k=K+1}^{\infty}Y_k+\sum_{k=-\infty}^{-K-1}Y_k.
		\label{eq:bdysumdecomposition}
	\end{equation}
	The last two terms are the positive and negative tails omitted by the cutoff.
	
	For a sufficiently decreasing function $f$, the Euler--Maclaurin formula gives
	\begin{equation}
		\sum_{k=K+1}^{\infty}f(k)=\int_K^\infty f(x)\,\dd x-\frac12f(K)+O\left(f'(K)\right).
		\label{eq:bdyEMtail}
	\end{equation}
	The endpoint term has a minus sign because the sum starts at $K+1$, whereas the integral starts at $K$.
	
	For the positive tail, use $f_+(x)=C_+x^{\rho-1}$. Since $\operatorname{Re}\rho<0$,
	\begin{equation}
		\int_K^\infty C_+x^{\rho-1}\,\dd x=-\frac{C_+K^\rho}{\rho}.
		\label{eq:bdypositiveintegral}
	\end{equation}
	Using Eq.~\eqref{eq:bdyYplus},
	\begin{equation}
		C_+K^\rho=KY_K+O(K^{\rho-1}).
		\label{eq:bdypositivepower}
	\end{equation}
	It follows that
	\begin{equation}
		\sum_{k=K+1}^{\infty}Y_k=-\left(\frac{K}{\rho}+\frac12\right)Y_K+O(K^{\rho-1}).
		\label{eq:bdypositivetail}
	\end{equation}
	
	For the negative tail, write $k=-\ell$ with $\ell>0$. Then
	\begin{equation}
		\sum_{k=-\infty}^{-K-1}Y_k=\sum_{\ell=K+1}^{\infty}Y_{-\ell}.
		\label{eq:bdynegativerewrite}
	\end{equation}
	Using Eq.~\eqref{eq:bdyYminus} in the same Euler--Maclaurin formula gives
	\begin{equation}
		\sum_{k=-\infty}^{-K-1}Y_k=-\left(\frac{K}{\rho}+\frac12\right)Y_{-K}+O(K^{\rho-1}).
		\label{eq:bdynegativetail}
	\end{equation}
	Therefore,
	\begin{equation}
		\sum_{|k|>K}Y_k=-\left(\frac{K}{\rho}+\frac12\right)\left(Y_K+Y_{-K}\right)+O(K^{\rho-1}).
		\label{eq:bdytail}
	\end{equation}
	
	The partial sum contains the opposite cutoff-dependent contribution. The finite-interval Euler--Maclaurin formula is
	\begin{equation}
		\sum_{k=1}^{K}f(k)=\int_1^Kf(x)\,\dd x+\frac{f(1)+f(K)}{2}+\frac{B_2}{2!}\left[f'(K)-f'(1)\right]+\cdots,
		\label{eq:bdyEMfinite}
	\end{equation}
	where $B_2=1/6$. Terms evaluated at the fixed lower endpoint are independent of $K$. Applying this formula to both asymptotic directions gives
	\begin{equation}
		\sum_{|k|\le K}Y_k=F_{\rm fin}+\left(\frac{K}{\rho}+\frac12\right)\left(Y_K+Y_{-K}\right)+O(K^{\rho-1}),
		\label{eq:bdypartial}
	\end{equation}
	where $F_{\rm fin}$ is independent of $K$.
	
	The cancellation can now be displayed directly:
	\begin{equation}
		\begin{split}
			F_\Lambda={}&\left[F_{\rm fin}+\left(\frac{K}{\rho}+\frac12\right)\left(Y_K+Y_{-K}\right)\right]
			\\
			&-\left(\frac{K}{\rho}+\frac12\right)\left(Y_K+Y_{-K}\right)+O(K^{\rho-1})
			\\
			={}&F_{\rm fin}+O(K^{\rho-1}).
		\end{split}
		\label{eq:bdyexplicitcancellation}
	\end{equation}
	The first term contains the cutoff dependence of the partial sum. The second term is the asymptotic contribution of the omitted tails. Their cutoff-dependent parts have opposite signs.
	
	Using
	\begin{equation}
		\frac1\rho=\frac{\Delta}{2i(n+1)\eta}=-\frac{i\Delta}{2(n+1)\eta},
		\label{eq:bdyinvrho}
	\end{equation}
	we obtain
	\begin{equation}
		-\left(\frac{K}{\rho}+\frac12\right)=\frac{iK\Delta}{2(n+1)\eta}-\frac12.
		\label{eq:bdytailcoefficient}
	\end{equation}
	The physical one-root amplitude is defined by
	\begin{equation}
		\begin{split}
			F_\Lambda^{\rm phys}(\bm z)={}&\lim_{\eta\to1+i0}\lim_{K\to\infty}\Biggl[\sum_{|k|\le K}Y_k
			\\
			&\quad+\left(\frac{iK\Delta}{2(n+1)\eta}-\frac12\right)\left(Y_K+Y_{-K}\right)\Biggr].
		\end{split}
		\label{eq:bdyregsum}
	\end{equation}
	
	For $\operatorname{Im}\eta>0$, the second term in Eq.~\eqref{eq:bdyregsum} is the Euler--Maclaurin approximation to the two omitted tails. At $\eta=1$, the individual terms decay only as $1/|k|$, so the physical answer is defined by analytic continuation from $\operatorname{Im}\eta>0$. The added term is a summation prescription. It is not an additional term in the Hamiltonian and does not modify the boundary reflection matrix.
	
	For several auxiliary roots, the same prescription is applied to the multiple lattice sum. The sum is first defined in the region $\operatorname{Im}\eta>0$, where it converges absolutely, and is then continued to $\eta\to1+i0$.
	
	The large-cutoff behavior also has a direct scattering interpretation. The singlet reflection eigenvalue in Eq.~\eqref{eq:ksinglet} satisfies
	\begin{equation}
		k(z)=1+\frac{4i\eta}{z}+O(z^{-2}),\qquad |z|\to\infty.
		\label{eq:bdysingletasymptotic}
	\end{equation}
	Successive round trips change the spectral label by $\Delta$, so the accumulated reflection phase contains
	\begin{equation}
		\sum_{p=1}^{K}\frac{4i\eta}{p\Delta}=\frac{4i\eta}{\Delta}\log K+O(1).
		\label{eq:bdylogphase}
	\end{equation}
	This logarithm is the phase counterpart of the power $K^\rho$ in the asymptotic tail. It arises because the bulk and boundary couplings decrease as the inverse of the spectral coordinate. The correction in Eq.~\eqref{eq:bdyregsum} removes the cutoff-dependent part of this asymptotic reflection history while retaining its finite contribution.
	
	\subsection{Extension beyond the explicit \texorpdfstring{$SU(2)$}{SU(2)} solution}
	\label{sec:su2extension}
	
	The $SU(2)$ calculation makes the general structure visible. The impurity is an additional quantum site, its spectral label is fixed, and in the round-trip transport each electron passes every other site twice, once before and once after the reflection, as in a double row. The Jackson sum works because shifting one electron label by $\Delta$ changes the weight by the scalar factors appearing in the off-shell unwanted terms. The terms cancel between neighboring points of the summation lattice.
	
	For $SU(N)$ we expect this argument to apply to the rational matrix
	\(S(u)=(iu\,\id+\PP)/(iu+1)\), with the nested monodromy matrix and its successive creation operators in place of the single operator $\bar{\mathsf B}$. The impurity is one extra spin site, so at the first level of nesting its parameter $d$ enters the weight as an additional site label. As in the periodic case, the coefficient of the first-level sum must then solve a boundary qKZ problem of lower rank for the roots. We have not carried out this nested construction here.
	
	Three extensions remain. Matrix boundaries of the form of Eq.~\eqref{eq:Kcnumbernorm} with commuting $M_\ell$ and $M_\rr$ can be diagonalized simultaneously and fall under the diagonal case of Refs.~\cite{ReshetikhinStokmanVlaar2015,ReshetikhinStokmanVlaar2018}. Noncommuting $M_\ell$ and $M_\rr$ break the $U(1)$ symmetry and require the off-diagonal Bethe ansatz~\cite{WangYangCaoShi2015}. The exchange-normalized reflection matrix of Sec.~\ref{sec:tracenorm} multiplies $K_\ell$ by the scalar $\sigma(u)$. This adds to every solution a scalar factor $\varpi(z_1)\cdots\varpi(z_n)$, where $\varpi(z+\Delta)=\sigma(z)\varpi(z)$, which can be written as a product over earlier reflections as in Eq.~\eqref{eq:Kprod}. The $O(3)$ realization of Sec.~\ref{sec:o3} has spin-$1$ bulk particles and a spin-$\tfrac12$ impurity. The same construction applies with Bethe vectors for tensor products of higher-spin modules~\cite{MukhinVarchenko1997,ReshetikhinStokmanVlaarFusion2016}, at the cost of weights with spin-dependent shifts.
	
	More generally, the construction should apply to other finite-dimensional representations whenever the mixed bulk--impurity scattering matrix and the endpoint reflection matrix satisfy the Yang--Baxter and reflection equations. In a fused or higher-spin representation, the scalar weights acquire representation-dependent shifts and the Bethe vectors change accordingly, while the affine label and the round-trip shift are unchanged. Whether the lattice sums then converge and span all solutions requires a separate analysis for each representation.
	
	\section{Small-step limit for the \texorpdfstring{$SU(2)$}{SU(2)} impurity}
	\label{sec:smallstep}
	
	This limit is taken for the rational $SU(2)$ bulk with the fundamental
	$SU(2)$ impurity solved above. The
	boundary qKZ equations reduce to the corresponding open-chain Bethe ansatz
	in the zero-step limit. The finite-difference transport becomes a
	double-row transfer-matrix eigenvalue problem, while the quasi-constant
	factor becomes the plane-wave phase.
	
	Take $\kappa\to0$ while keeping the physical momenta fixed, and use the
	plane-wave quasi-constant
	\begin{equation}
		F(\bm z)\sim
		\exp\!\left[-\frac{i}{\kappa}\sum_{j=1}^{n}k_jz_j\right]
		\Xi(\bm z).
	\end{equation}
	Since $\Delta/\kappa=2\sL$, Eq.~\eqref{eq:BqKZ} becomes, to leading order in $\kappa$,
	\begin{equation}
		e^{-2ik_j\sL}\,\Xi
		=\mathcal T_j^{(0)}\,\Xi .
		\label{eq:smallsteptransfer}
	\end{equation}
	Here $\mathcal T_j^{(0)}$ is the transport~\eqref{eq:doubleRow} at $\Delta=0$ with every electron label equal to
	\begin{equation}
		b=u_{B,0},
	\end{equation}
	because each label $z_j=\kappa(x_j+t)+u_{B,0}$ tends to $u_{B,0}$ as $\kappa\to0$, while the impurity keeps its label $d$. The same-chirality factors then reduce to $S(0)=\PP$ and the right--left factors to the static bulk matrix $S(2b)$. Thus $\mathcal T_j^{(0)}$ is the round-trip transport of a static open rational $SU(2)$ chain with $n$ sites of label $b$, one site of label $d$, and trivial auxiliary boundary factors. It is proportional to Sklyanin's transfer matrix. For arbitrary labels $z_0,\ldots,z_n$ at $\Delta=0$, the monodromy matrix of Eq.~\eqref{eq:doublerow} gives
	\begin{equation}
		\mathcal T_j^{(0)}=\frac{2z_j-i\eta}{2(z_j-i\eta)}\,\Tr_a\,\mathbb U_a(-z_j),
		\label{eq:Ttransfer}
	\end{equation}
	because at $\theta=-z_j$ the factor $S_{aj}(0)=\PP_{aj}$ turns the trace over the auxiliary particle into the transport of particle $j$. The trace closes the double row with the identity, which is the dual reflection matrix corresponding to $K_\rr=\id$. This $K^+$ belongs only to the comparison with Sklyanin's transfer matrix; it is not an additional factor in the boundary transport~\eqref{eq:doubleRow}, which contains the physical matrix $K_\rr$ itself. For a nontrivial right reflection of the form~\eqref{eq:Kcnumbernorm}, the same comparison at $\Delta=0$ holds with the trace closed by
	\begin{equation}
		K^+(\theta)=K_\rr(\theta+i\eta),
		\label{eq:staticdual}
	\end{equation}
	the crossing dual of $K_\rr$, whose traces commute (App.~\ref{app:bqkzproof}). At $\Delta\ne0$ and $\xi_\rr\ne0$ the matrix required at the far end depends on $\Delta$ and its traces do not commute, so the finite-step transport with a nontrivial $K_\rr$ is not a Sklyanin transfer matrix. Only in the static limit do the two coincide. Equation~\eqref{eq:Ttransfer} follows algebraically from $S_{aj}(0)=\PP_{aj}$, the Yang--Baxter equation, and braiding unitarity. Its prefactor is the inverse of $\Tr_a[S_{aj}(2z_j)\PP_{aj}]=2(z_j-i\eta)/(2z_j-i\eta)$. The identity~\eqref{eq:Ttransfer} is the precise sense in
	which the small-step limit gives Bethe quantization. The limit is not
	obtained by setting the scattering matrix to the identity; it is a
	simultaneous zero-step and rapidly oscillating-wave limit.
	
	The eigenvalues of $\mathcal T_j^{(0)}$ follow from the Jackson sums of Sec.~\ref{sec:kondosolution} without further input. As $\Delta\to0$ the lattice spacing vanishes, and the sum over $\Lambda$ is dominated by roots at which the weight $\Phi(\bm z,\bm\theta)$ of Eq.~\eqref{eq:bdyPhi} is stationary under a shift of one root by $\Delta$. At such roots the unwanted terms vanish separately instead of canceling between neighboring lattice points, and $\Omega(\bm z;\bm\theta)$ becomes an eigenvector of $\mathcal T_j^{(0)}$. The shift relations~\eqref{eq:bdyWshift} and~\eqref{eq:bdyGshift} are exact, and at fixed $\bm z$ and $\bm\theta$ they give
	\begin{equation}
		\begin{split}
			\lim_{\Delta\to0}\frac{\Phi(\bm z,\bm\theta+\Delta e_m)}{\Phi(\bm z,\bm\theta)}
			={}&\prod_{r=0}^{n}\prod_{\upsilon=\pm}\frac{\theta_m+\upsilon z_r+i\eta/2}{\theta_m+\upsilon z_r-i\eta/2}\\
			&\times\prod_{m'\ne m}\prod_{\upsilon=\pm}\frac{\theta_m+\upsilon\theta_{m'}-i\eta}{\theta_m+\upsilon\theta_{m'}+i\eta},
		\end{split}
		\label{eq:rootratio}
	\end{equation}
	\begin{equation}
		\lim_{\Delta\to0}\frac{\Phi(\bm z+\Delta e_j,\bm\theta)}{\Phi(\bm z,\bm\theta)}
		=\prod_{m=1}^{M}\prod_{\upsilon=\pm}\frac{z_j+\upsilon\theta_m+i\eta/2}{z_j+\upsilon\theta_m-i\eta/2}.
		\label{eq:labelratio}
	\end{equation}
	The first ratio contains the factors $\mathcal W(z_r\pm\theta_m)$ and $\mathcal G(\theta_m\pm\theta_{m'})$, and the second the factors $\mathcal W(z_j\pm\theta_m)$. We set $z_0=d$, $z_1=\cdots=z_n=b$, and call the stationary roots $\lambda_\alpha$. Stationarity sets the ratio~\eqref{eq:rootratio} equal to one, and the ratio~\eqref{eq:labelratio} is the eigenvalue of $\mathcal T_j^{(0)}$, which by Eq.~\eqref{eq:smallsteptransfer} equals $e^{-2ik_j\sL}$. This gives
	\begin{equation}
		e^{-2ik_j\sL}
		=\prod_{\alpha=1}^{M}\prod_{\upsilon=\pm}
		\frac{b+\upsilon\lambda_\alpha+i\eta/2}
		{b+\upsilon\lambda_\alpha-i\eta/2},
		\qquad j=1,\ldots,n,
		\label{eq:staticchargeBA}
	\end{equation}
	and
	\begin{equation}
		\begin{split}
			&\prod_{\upsilon=\pm}
			\left(\frac{\lambda_\alpha+\upsilon b+i\eta/2}
			{\lambda_\alpha+\upsilon b-i\eta/2}\right)^{n}
			\left(\frac{\lambda_\alpha+\upsilon d+i\eta/2}
			{\lambda_\alpha+\upsilon d-i\eta/2}\right)\\
			&\qquad=\prod_{\upsilon=\pm}\prod_{\substack{\beta=1\\\beta\ne\alpha}}^{M}
			\frac{\lambda_\alpha+\upsilon\lambda_\beta+i\eta}
			{\lambda_\alpha+\upsilon\lambda_\beta-i\eta},
			\qquad \alpha=1,\ldots,M .
		\end{split}
		\label{eq:staticspinBA}
	\end{equation}
	These are the Bethe equations of the open rational chain with $n$ sites of label $b$, one site of label $d$, and trivial reflection and dual reflection matrices~\cite{Sklyanin1988}, in agreement with Eq.~\eqref{eq:Ttransfer}. The factor containing $d$ is the impurity contribution, and the factor containing $b$ is the contribution of the $n$ electrons. The total spin is $S^z=(n+1)/2-M$, because the impurity contributes one additional spin-$\tfrac12$ site. The stationarity argument is the boundary counterpart of the small-step limit of Jackson integrals for the periodic qKZ equations~\cite{reshetikhin1992jackson}.
	
	The Bethe equations contain only the labels $b$ and $d$, because the boundary qKZ system contains only $S$ and $K_\ell$. Their relation to the couplings of the static Hamiltonian does not come from the small-step limit. It follows from the reconstruction of Sec.~\ref{sec:cayley}, evaluated at $\kappa=0$, where $u_B(t)=b$ at all times. Equation~\eqref{eq:HbulkSUN} then has the coefficient $g=1/(2b)$ multiplying $\PP-\id$, and Eq.~\eqref{eq:HKondoboundary} has the coefficient
	\begin{equation}
		c=\frac{2b}{D(b)}=\frac{2b}{b^2-1-d^2}
		\label{eq:cSU2}
	\end{equation}
	multiplying $\PP_{a0}-\id$. Solving Eq.~\eqref{eq:cSU2} for the impurity label gives
	\begin{equation}
		d^2=b^2-\frac{2b}{c}-1 .
		\label{eq:dSU2}
	\end{equation}
	The constant $1$ is $\eta^2$ at $\eta=1$ and comes from the shifts $\pm i$ in the impurity reflection matrix, Eq.~\eqref{eq:kSUNimpA}. The $O(3)$ impurity has the shifts $\pm3i/2$, which give the constant $\tfrac94$ in Eq.~\eqref{eq:dRG}. Equation~\eqref{eq:dSU2} does not involve the bulk coefficient $g$, so it holds for every value of $b$. The label $d$ is real or purely imaginary, as allowed in Sec.~\ref{sec:impurity}.
	
	With the exchange normalization of Sec.~\ref{sec:tracenorm}, $\mathcal T_j^{(0)}$ is multiplied by the scalar $\sigma(b)$. This shifts all momenta in Eq.~\eqref{eq:staticchargeBA} by the same amount and leaves Eq.~\eqref{eq:staticspinBA} unchanged. We write the exchange-normalized interaction matrix as
	\begin{equation}
		B^{\rm exch}_{a0}(b;d)=J\,t_a^A\tau_0^A,
	\end{equation}
	where $\tau_0^A$ are the Pauli matrices of the impurity, so that $g_\partial(b)=4J$ in Eq.~\eqref{eq:BexchSUN}. On the triplet and the singlet of the electron and the impurity, $t^A\tau_0^A$ has the eigenvalues $\tfrac12$ and $-\tfrac32$, so the Cayley parameters of $K^{\rm exch}_\ell$ are $J/2$ and $-3J/2$. The first is the parameter $a_\sigma$ of the scalar phase. The second combines $a_\sigma$ with the singlet parameter $a_-=-2b/D(b)=-c$ of $K_\ell$ according to Eq.~\eqref{eq:tanadd},
	\begin{equation}
		\frac{J/2-c}{1+cJ/2}=-\frac{3J}{2}.
	\end{equation}
	Hence
	\begin{equation}
		c=\frac{2J}{1-3J^2/4},
		\label{eq:cJSU2}
	\end{equation}
	the $SU(2)$ counterpart of Eq.~\eqref{eq:cJ}. Both roots of the quadratic~\eqref{eq:bdyquadratic} satisfy Eq.~\eqref{eq:cJSU2}, so the relation does not depend on the branch chosen in Sec.~\ref{sec:tracenorm}. Equations~\eqref{eq:dSU2} and~\eqref{eq:cJSU2} express the impurity label through the static exchange coupling in the matching convention of App.~\ref{app:transmission}. A different short-distance convention for the contact interactions would change these relations, but not the Bethe equations~\eqref{eq:staticchargeBA} and~\eqref{eq:staticspinBA}.
	
	\section{\texorpdfstring{$O(3)$}{O(3)}-vector bulk and
		\texorpdfstring{$SU(2)$}{SU(2)}-doublet impurity}
	\label{sec:o3}
	
	\subsection{Bulk and impurity representation spaces}
	
	The bulk color space is $\cV=\mathbb C^3$ with generators $(\tau^a)_{bc}=-i\epsilon_{abc}$, while the impurity space is $\cBleft=\mathbb C^2$ with Pauli matrices $\sigma^a$. The bulk particle and the impurity thus carry the spin-$1$ and spin-$\tfrac12$ representations of the same algebra $\mathfrak{su}(2)\simeq\mathfrak{so}(3)$. This allows the bulk and impurity degrees of freedom to couple through the invariant $\vec\tau\cdot\vec\sigma$. 
	
	\subsection{\texorpdfstring{$O(3)$}{O(3)} invariant bulk scattering}
	
	Define
	\begin{equation}
		(\PP_{12})_{mn}{}^{pq}=\delta_m^q\delta_n^p, \qquad (\Q_{12})_{mn}{}^{pq}=\delta_{mn}\delta^{pq},
		\label{eq:PC}
	\end{equation}
	satisfying $\PP^2=\id$, $\Q^2=3\Q$, $\PP\Q=\Q\PP=\Q$. Both are invariant under conjugation by the permutation, so the symmetry $S_{21}(u)=S_{12}(u)$ holds, and the algebra they generate is commutative. From $\tau^a_{bc}\tau^a_{de}=-\delta_{bd}\delta_{ce}+\delta_{be}\delta_{cd}$ one finds $\vec\tau_1\cdot\vec\tau_2=\PP_{12}-\Q_{12}$. The irreducible projectors are
	\begin{equation}
		\mathcal P^{(0)}=\tfrac13\Q, \qquad \mathcal P^{(1)}=\tfrac12(\id-\PP),
	\end{equation}
	\begin{equation}
		\mathcal P^{(2)}=\tfrac12(\id+\PP)-\tfrac13\Q,
		\label{eq:O3projectors}
	\end{equation}
	of ranks $1,3,5$. On them $(\PP,\Q)$ takes the values $(1,3)$, $(-1,0)$, $(1,0)$, so that $\vec\tau_1\cdot\vec\tau_2$ takes $-2,-1,+1$, in agreement with $\tfrac12[J(J+1)-4]$.
	
	The normalized bulk matrix is
	\begin{equation}
		S_{12}^{O(3)}(u)=s_0(u)\mathcal P^{(0)}+s_1(u)\mathcal P^{(1)} +\mathcal P^{(2)},
		\label{eq:SO3projector}
	\end{equation}
	\begin{equation}
		s_0(u)=\frac{(u+i)(u+2i)}{(u-i)(u-2i)}, \qquad s_1(u)=\frac{u+2i}{u-2i},
		\label{eq:O3eigen}
	\end{equation}
	unitary for real $u$ and satisfying $S^{O(3)}(u)S^{O(3)}(-u)=\id$. We normalize the $\mathcal P^{(2)}$ channel to unity. The remaining eigenvalue ratios agree with those of the Zamolodchikov $O(3)$ $S$-matrix~\cite{ZZ1979}. We first record this rational representative. Since $s_0(0)=1$ and $s_1(0)=-1$,
	\begin{equation}
		S^{O(3)}(0)=\mathcal P^{(0)}-\mathcal P^{(1)}+\mathcal P^{(2)}=\PP_{12},
		\label{eq:O3Szero}
	\end{equation}
	for the rational normalization. Equivalently
	\begin{equation}
		S_{12}^{O(3)}(u)=\frac{s_1+1}{2}\id +\frac{1-s_1}{2}\PP +\frac{s_0-1}{3}\Q ,
		\label{eq:SO3IPC}
	\end{equation}
	which has the same operator structure as the $19$-vertex $R$-matrix used in Ref.~\cite{KattelO3SC2025} for the integrable $O(3)$-invariant Gross--Neveu-like model of Ref.~\cite{AndreiDestri1984}.
	
	Applying the bulk inversion formula \eqref{eq:bulkCayley} to the rational representative gives
	\begin{equation}
		V_{12}^{O(3)}(u) =-\frac{1}{u}\bigl(\id-\PP_{12}\bigr)-\frac{u}{u^2-2}\,\Q_{12},
		\label{eq:VO3}
	\end{equation}
	with eigenvalues $0$, $-2/u$, and $-3u/(u^2-2)$ on $\mathcal P^{(2)}$, $\mathcal P^{(1)}$, and $\mathcal P^{(0)}$. We do not use this rational representative for the driven Hamiltonian. We instead choose a scalar normalization whose Cayley inverse has the current--current form. Define
	\begin{equation}
		\rho_{O(3)}(u)=\frac{u-2i}{u+2i}=s_1(u)^{-1},
		\label{eq:O3currentphase}
	\end{equation}
	with $\widetilde S^{O(3)}(u)=\rho_{O(3)}(u)S^{O(3)}(u)$.
	Applying Eq.~\eqref{eq:bulkCayley} to $\widetilde S^{O(3)}$ gives Cayley eigenvalues $2/u$, $0$, and $-1/u$ on the $J=2,1,0$ channels. Since $\id+\vec\tau_1\cdot\vec\tau_2$ has eigenvalues $2,0,-1$ on the same channels,
	\begin{equation}
		\widetilde V_{12}^{O(3)}(u)=\frac{1}{u}\bigl(\id+\vec\tau_1\cdot\vec\tau_2\bigr).
		\label{eq:VO3current}
	\end{equation}
	The phase obeys $\rho_{O(3)}(u)\rho_{O(3)}(-u)=1$, so this normalization preserves spectral unitarity and the BqKZ scalar condition. We use this representative for the $O(3)$ bulk matching, round-trip transport, and Hamiltonian below. Evaluating Eq.~\eqref{eq:VO3current} at the physical argument $2u_B(t)$ gives the time-dependent bulk coupling directly.
	
	\subsection{Mixed bulk--impurity scattering}
	\label{sec:mixed}
	
	Let $Y_{a0}=\vec\tau_a\cdot\vec\sigma_0$. Since $\vec\tau\cdot\vec\sigma=2\vec\tau\cdot\vec s$ and $2\vec\tau\cdot\vec s=J(J+1)-2-\tfrac34$, the eigenvalues are $+1$ on the quartet and $-2$ on the doublet, with projectors
	\begin{equation}
		\mathcal P_{\frac12}=\tfrac13(\id-Y_{a0}), \qquad \mathcal P_{\frac32}=\tfrac13(2\id+Y_{a0}).
		\label{eq:mixedprojectors}
	\end{equation}
	The normalized mixed matrix is
	\begin{equation}
		s_{a0}(u)=\frac{\bigl(u-\tfrac{i}{2}\bigr)\id_{a0}-iY_{a0}} {u-\tfrac{3i}{2}},
		\label{eq:mixedO3}
	\end{equation}
	equal to $\id$ on the quartet and to $(u+\tfrac{3i}{2})/(u-\tfrac{3i}{2})$ on the doublet. It is unitary for real $u$, satisfies $s_{a0}(u)s_{a0}(-u)=\id$, and commutes at different spectral arguments because it is a function of $Y_{a0}$ alone.
	
	The bulk and mixed matrices use the same spectral normalization; App.~\ref{app:o3spectral} gives the explicit comparison.
	
	We place the impurity at the left endpoint and take
	\begin{equation}
		K_{\ell,a0}^{O(3)}(u;d)=s_{a0}(u+d)\,s_{a0}(u-d),
		\label{eq:KO3imp}
	\end{equation}
	which solves the left reflection equation. The matrix $K^{O(3)}_{\ell,a0}$ has quartet eigenvalue $1$ and doublet eigenvalue
	\begin{equation}
		\lambda(u;d)=\frac{(u+d+\tfrac{3i}{2})(u-d+\tfrac{3i}{2})} {(u+d-\tfrac{3i}{2})(u-d-\tfrac{3i}{2})} =\frac{E(u)+3iu}{E(u)-3iu},
		\label{eq:lambdaO3}
	\end{equation}
	\begin{equation}
		E(u)=u^2-d^2-\tfrac94 ,
		\label{eq:Edef}
	\end{equation}
	using $(u\pm\tfrac{3i}2)^2-d^2=E\pm3iu$. Hence $|\lambda|=1$ whenever $d^2\in\mathbb R$. We take the right endpoint to be $K_\rr=\id$.
	
	For the bulk factorization of this example, we use the same representative as in the Hamiltonian, $S^{RL}=S^{LR}=\widetilde S^{O(3)}$ and $\mathcal S^{RR}_{jk}=\mathcal S^{LL}_{jk}=\widetilde S^{O(3)}_{jk}(\zeta_j-\zeta_k)$. The scalar phase obeys the BqKZ condition, so this replacement preserves the round-trip flatness relation. Same-chirality factors remain auxiliary and contribute no local interaction.
	
	\subsection{Boundary inversion and the driven Kondo coupling}
	\label{sec:o3check}
	
	Applying Eq.~\eqref{eq:boundary-inverse} to Eq.~\eqref{eq:KO3imp} channel by channel gives
	\begin{equation}
		B_{a0}^{O(3)}(u;d)=-\frac{3u}{E(u)}\,\mathcal P_{\frac12}
		=-\frac{u}{E(u)}\bigl(\id-Y_{a0}\bigr).
		\label{eq:BO3imp}
	\end{equation}
	This rational interaction contains both a scalar part and the $O(3)$-covariant exchange. As in the $SU(N)$ impurity, we multiply the reflection matrix by a scalar phase,
	\begin{equation}
		K_{\ell,a0}^{O(3),\rm exch}(u;d)=\sigma_{O(3)}(u)K_{\ell,a0}^{O(3)}(u;d),
		\label{eq:KO3exch}
	\end{equation}
	and choose the phase so that the identity component of the Cayley inverse vanishes. Applying Eq.~\eqref{eq:boundary-inverse} to Eq.~\eqref{eq:KO3exch} gives
	\begin{equation}
		\begin{split}
			B_{a0}^{\rm exch}(u;d)&=J(u;d)Y_{a0},\\
			J(u;d)&=\frac{\sqrt{E(u)^2+8u^2}-E(u)}{4u}.
		\end{split}
		\label{eq:JO3u}
	\end{equation}
	App.~\ref{app:normalization} derives the scalar phase and branch choice. The characteristic geometry fixes the physical argument to $u=u_\ell(t)=u_B(t)$, so
	\begin{equation}
		J(t)=\frac{\sqrt{E(u_B(t))^2+8u_B(t)^2}-E(u_B(t))}{4u_B(t)}.
		\label{eq:JO3t}
	\end{equation}
	Substituting Eq.~\eqref{eq:JO3u} into the general impurity Hamiltonian \eqref{eq:Hboundary} gives
	\begin{equation}
		H_K(t)=2J(t)\,
		\bigl(\psi_R^\dagger\vec\tau\,\psi_L\bigr)(0,t)\cdot\vec\sigma_0.
		\label{eq:HO3exch}
	\end{equation}
	App.~\ref{app:cayleydomain} collects the exceptional values of this local parametrization.
	
	\subsection{The driven \texorpdfstring{$O(3)$}{O(3)} Hamiltonian}
	
	For the bulk, we use the BqKZ-compatible current--current representative \eqref{eq:VO3current}. Inserting its physical argument $u=2u_B(t)$, as prescribed by the general bulk Hamiltonian \eqref{eq:Hbulk}, gives
	\begin{equation}
		\widetilde V^{O(3)}\bigl(2u_B(t)\bigr)
		=\frac{1}{2u_B(t)}
		\bigl(\id+\vec\tau_1\cdot\vec\tau_2\bigr).
		\label{eq:VO3physical}
	\end{equation}
	It is convenient to write
	\begin{equation}
		g(t)=\frac{1}{2u_B(t)},
		\label{eq:gO3t}
	\end{equation}
	so that the coefficient in the second-quantized Hamiltonian is $2g(t)=1/u_B(t)$. Combining the bulk and impurity interactions gives
	\begin{equation}
		\begin{split}
			H_{O(3)}(t)={}&H_0
			+2g(t)\int_0^{\sL}\dd x\;
			:\psi_R^\dagger\psi_L^\dagger
			\bigl(\id+\vec\tau_1\cdot\vec\tau_2\bigr)
			\psi_L\psi_R:\\
			&+2J(t)\,
			\bigl(\psi_R^\dagger\vec\tau\,\psi_L\bigr)(0,t)
			\cdot\vec\sigma_0,
		\end{split}
		\label{eq:HO3total}
	\end{equation}
	with $u_B(t)=u_{B,0}+\kappa t$, $g(t)$ given by Eq.~\eqref{eq:gO3t}, and $J(t)$ by Eq.~\eqref{eq:JO3t}. We take the right endpoint to be trivial. Thus $u_B(t)$ fixes both the bulk coupling $g(t)$ and the Kondo coupling $J(t)$, while $d$ remains a fixed impurity parameter.
	
	\section{Conclusion and outlook}
	\label{sec:conclusion}
	
	We developed a general construction of time-dependent integrable chiral quantum field theories on a finite interval starting from autonomous bulk and boundary scattering data. Given a bulk $S$-matrix and compatible reflection matrices $K_\ell$ and $K_\rr$, spatial homogeneity fixes the spectral labels carried by the chiral particles to be affine functions of their characteristics. The same affine coordinate then fixes the arguments of the bulk and boundary matrices. In our conventions, a right--left scattering event samples $S(2u_B(t))$, while the two endpoints sample $K_\ell(u_B(t))$ and $K_\rr(-u_B(t))$. The time dependence of the bulk and boundary interactions is determined by the scattering and reflection data together with the characteristic map rather than introduced independently.
	
	On a finite interval, local factorization alone is not sufficient. Successive reflections at the two endpoints translate the spectral label by $\Delta=2\kappa\sL$, and following a particle through a complete round trip gives a difference transport in spectral space. The Yang--Baxter and reflection equations ensure consistency of the local scattering and reflection processes, while boundary qKZ flatness ensures consistency of the complete scattering and reflection circuits. These relations provide the local and global conditions for the time-dependent factorized evolution.
	
	The construction reduces the many-body problem to the boundary qKZ system. A solution of these equations determines the coefficient functions in spectral space, and the characteristic pullback gives the corresponding time-dependent wavefunction in each finite-particle sector above the bare Fock vacuum. Independently, inverting the scattering and reflection matrices through the local matching problem gives the local Hamiltonian. Thus the autonomous $S$- and $K$-matrices determine the boundary qKZ transport problem and the allowed local bulk and boundary interactions, while a solution of the qKZ system determines the corresponding factorized wavefunction.
	
	The examples show how these general statements work in practice. For rational $SU(N)$ scattering, we obtained a driven bulk contact interaction together with two time-dependent matrix boundaries, and a fundamental boundary impurity gives a driven $SU(N)$ Kondo interaction. For the $O(3)$ vector model with a spin-$\tfrac12$ boundary impurity, this construction gives a time-dependent current--current bulk interaction and a driven Kondo coupling. In each case, the allowed time dependence follows from the affine coordinate $u_B(t)$, while the remaining parameters specify the autonomous scattering or boundary data.
	
	The main result is not tied to these examples. Starting from suitable bulk and boundary factorized data, the construction provides a route to a time-dependent integrable model on a finite interval, determines the allowed bulk and boundary drive, and reduces its exact finite-particle dynamics to a boundary qKZ problem. This opens the construction to other $R$-matrices, reflection matrices, representations, and boundary quantum degrees of freedom. For the $SU(2)$ Kondo impurity we constructed explicit solutions of the boundary qKZ system, in which the impurity acts as an additional site with a fixed label and the roots are summed over a lattice together with their mirror images. For one electron the singlet amplitude is a ratio of Gamma functions that collects the phases of all earlier reflections at the impurity. When the drive is switched off, the boundary qKZ system reduces to the Bethe equations of the static open chain, with the impurity as an inhomogeneity. It would be interesting to extend these solutions to the $O(3)$ impurity, to noncommuting matrix boundaries, and to dressed many-body reference states beyond the bare Fock vacuum.

	\begin{acknowledgments}
		This work is supported by the Swiss National Science Foundation under grant number 200020-219400. OpenAI
		ChatGPT (GPT-5.6 Luna) was used to polish the final document drafted by the author.
	\end{acknowledgments}
	
	\appendix
	
	\section{Local matching and Cayley inversion}
	\label{app:transmission}
	
	The one- and two-particle matching convention of the microscopic chiral Hamiltonian defines the local couplings used in the main text. For a local matrix $X$ the corresponding unitary matching matrix is
	\begin{equation}
		U=(\id+iX)^{-1}(\id-iX),
		\label{eq:generic-forward}
	\end{equation}
	so that
	\begin{equation}
		X=i(U-\id)(U+\id)^{-1}.
		\label{eq:generic-inverse}
	\end{equation}
	This is the Cayley transform used for both the bulk and the impurity. In the bulk, $U=S$ and $X=V$. At an impurity, $U=K$ and $X=B$. The second-quantized interaction carries the conventional factor $2$ used in Eqs.~\eqref{eq:Hbulk} and \eqref{eq:Hboundary}. The relation between a bare coupling and a scattering parameter depends on the short-distance convention used to define the local interaction. Here we keep one convention throughout, fixed by the microscopic matching problem.
	
	On orthogonal projectors, $U=\sum_\mu\lambda_\mu\Pi_\mu$ gives $X=\sum_\mu x_\mu\Pi_\mu$ with
	\begin{equation}
		x_\mu=i\frac{\lambda_\mu-1}{\lambda_\mu+1}.
		\label{eq:channelrule}
	\end{equation}
	For a unitary matching matrix these eigenvalues are real wherever $-1\notin\Spec U$. It is convenient to write
	\begin{equation}
		\lambda_\mu=\frac{1-ia_\mu}{1+ia_\mu},\qquad x_\mu=a_\mu,
		\label{eq:tanparam}
	\end{equation}
	in the convention of Eq.~\eqref{eq:generic-inverse}. Multiplying two unimodular eigenvalues combines their Cayley parameters according to
	\begin{equation}
		a_1\oplus a_2=\frac{a_1+a_2}{1-a_1a_2}.
		\label{eq:tanadd}
	\end{equation}
	
	If an eigenvalue of $U$ reaches $-1$, the inverse formula for $X$ becomes infinite although $U$ itself remains finite. App.~\ref{app:cayleydomain} collects the examples in which this occurs.
	
	\section{The matrix boundary solves the reflection equation}
	\label{app:matrixRE}
	
	We verify Eq.~\eqref{eq:Kcnumbernorm} for the rational bulk matrix \eqref{eq:SUN}. Scalar prefactors cancel from the reflection equation, so it is enough to use
	\begin{equation}
		R(u)=iu\,\id+\PP_{12},\qquad \kappa_\alpha(u)=\xi_\alpha\id_N+iu\,M_\alpha.
		\label{eq:algrep}
	\end{equation}
	Write $x=iu$ and $y=iv$ and suppress the endpoint label. The reflection equation becomes
	\begin{equation}
		\begin{split}
			&R(x-y)\,\kappa_1(x)\,R(x+y)\,\kappa_2(y)\\
			&\qquad=\kappa_2(y)\,R(x+y)\,\kappa_1(x)\,R(x-y),
		\end{split}
		\label{eq:REalg}
	\end{equation}
	with $R(x)=x\id+\PP$.
	
	Use $\PP^2=\id$, $\PP M_1=M_2\PP$, and $\PP M_2=M_1\PP$. Expanding both sides, all terms proportional to $\xi^2$ and $\xi$ agree. Among the terms independent of $\xi$, the terms proportional to $(x^2-y^2)xyM_1M_2$, $(x+y)xyM_1M_2\PP$, and the terms without $\PP$ containing $M_1^2$ or $M_2^2$ agree as well. The only remaining comparison is between $xy(x-y)M_1^2\PP$ on one side and $xy(x-y)M_2^2\PP$ on the other. Thus the reflection equation requires $M_1^2\PP=M_2^2\PP$, which is satisfied whenever $M^2=c\,\id_N$. We can absorb the positive constant $c$ into a rescaling of $M$ together with $\xi$. We use the normalized form $M^2=\id_N$ in Eq.~\eqref{eq:Kcnumbernorm}.
	
	The mixed products that appear in the expansion are $M_1\PP M_2$ and $M_2\PP M_1$. There is no additional commutator condition, since $M_1$ and $M_2$ already act on different tensor factors. This completes the check of Eq.~\eqref{eq:RE}. The same family with $u\mapsto-u$ gives the opposite endpoint orientation used in Eq.~\eqref{eq:REright}.
	
	\section{Two-particle check of the flatness relation}
	\label{app:n2}
	
	For $n=2$ we can display the cancellations underlying Eq.~\eqref{eq:BqKZflat} in a few lines while keeping both endpoints arbitrary. The calculation uses the symmetry $S_{21}(u)=S_{12}(u)$, the braiding unitarity relation Eq.~\eqref{eq:Sunitarity}, and the reflection equation at each endpoint. We do not need the Yang--Baxter equation for $n=2$. It first enters at $n=3$.
	
	Write $\alpha=z_1-z_2$ and $\beta=z_1+z_2$. Eqs.
	\eqref{eq:T1n2}--\eqref{eq:T2n2} read
	\begin{equation}
		\begin{split}
			\mathcal T_1(z_1,z_2)&=S(\alpha+\Delta)\,K_{\rr,1}\left(-z_1-\tfrac{\Delta}{2}\right)\,S(\beta)\,K_{\ell,1}(z_1),\\
			\mathcal T_2(z_1,z_2)&=K_{\rr,2}\left(-z_2-\tfrac{\Delta}{2}\right)\,S(\beta)\,K_{\ell,2}(z_2)\,S(-\alpha).
		\end{split}
		\label{eq:n2transportcheck}
	\end{equation}
	
	Transport particle $1$ first. Under $z_1\to z_1+\Delta$ one has $\alpha\to \alpha+\Delta$ and $\beta\to \beta+\Delta$, so
	\begin{equation}
		\begin{split}
			&\mathcal T_2(z_1+\Delta,z_2)\,\mathcal T_1(z_1,z_2)\\
			&=K_{\rr,2}\left(-z_2-\tfrac{\Delta}{2}\right)S(\beta{+}\Delta)K_{\ell,2}(z_2)\,S(-\alpha{-}\Delta)S(\alpha{+}\Delta)\\
			&\qquad\times K_{\rr,1}\left(-z_1-\tfrac{\Delta}{2}\right)S(\beta)K_{\ell,1}(z_1).
		\end{split}
		\label{eq:n2sideA}
	\end{equation}
	The backtracking pair cancels, $S(-\alpha-\Delta)S(\alpha+\Delta)=\id$, by the symmetry $S_{21}(u)=S_{12}(u)$ and Eq.~\eqref{eq:Sunitarity}. The two factors that then become adjacent, $K_{\ell,2}(z_2)$ and $K_{\rr,1}\left(-z_1-\tfrac{\Delta}{2}\right)$, act on $\cV_2\otimes\cBleft$ and on $\cV_1\otimes\cBright$ respectively. They have no tensor factor in common and commute. Hence
	\begin{align}
		&K_{\rr,2}\left(-z_2-\tfrac{\Delta}{2}\right)\,S(\beta{+}\Delta)\,K_{\rr,1}\left(-z_1-\tfrac{\Delta}{2}\right)\; \nonumber\\
		&\qquad\times \Bigl[K_{\ell,2}(z_2)\,S(\beta)\,K_{\ell,1}(z_1)\Bigr].
		\label{eq:n2flatness}
	\end{align}
	
	Transport particle $2$ first. Under $z_2\to z_2+\Delta$ one has $\alpha\to \alpha-\Delta$ and $\beta\to \beta+\Delta$, so the leading factor of $\mathcal T_1$ becomes $S(\alpha)$, and commuting $K_{\ell,1}(z_1)$ past $K_{\rr,2}\left(-z_2-\tfrac{\Delta}{2}\right)$ for the same reason,
	\begin{equation}
		\begin{split}
			&\mathcal T_1(z_1,z_2+\Delta)\,\mathcal T_2(z_1,z_2)\\
			&=S(\alpha)\,K_{\rr,1}\left(-z_1-\tfrac{\Delta}{2}\right)\,S(\beta{+}\Delta)\,K_{\rr,2}\left(-z_2-\tfrac{\Delta}{2}\right)\\
			&\qquad\times\Bigl[K_{\ell,1}(z_1)\,S(\beta)\,K_{\ell,2}(z_2)\Bigr]S(-\alpha).
		\end{split}
		\label{eq:n2sideB}
	\end{equation}
	
	The two bracketed groups are related by the reflection equation at the left endpoint. Eq.~\eqref{eq:RE} with $u=z_1$ and $v=z_2$, so that $u-v=\alpha$ and $u+v=\beta$, gives
	\begin{equation}
		K_{\ell,2}(z_2)S(\beta)K_{\ell,1}(z_1) =S(\alpha)\,K_{\ell,1}(z_1)S(\beta)K_{\ell,2}(z_2)\,S(-\alpha)
	\end{equation}
	after multiplying by $S(-\alpha)$ on the right. Substituting this into Eq.~\eqref{eq:n2flatness}, the two orderings agree if and only if the right-endpoint factors obey
	\begin{equation}
		\begin{split}
			&K_{\rr,2}\left(-z_2-\tfrac{\Delta}{2}\right)\,S(\beta+\Delta)\,K_{\rr,1}\left(-z_1-\tfrac{\Delta}{2}\right)\,S(\alpha)\\
			&\qquad=S(\alpha)\,K_{\rr,1}\left(-z_1-\tfrac{\Delta}{2}\right)\,S(\beta+\Delta)\,K_{\rr,2}\left(-z_2-\tfrac{\Delta}{2}\right).
		\end{split}
		\label{eq:n2right}
	\end{equation}
	
	Eq.~\eqref{eq:n2right} is the reflection equation of the right endpoint. In the coordinate centered there the two incoming labels are $p=-z_1-\tfrac\Delta2$ and $q=-z_2-\tfrac\Delta2$ by Eq.~\eqref{eq:Krfactor}, and in terms of them $\alpha=q-p$ and $\beta+\Delta=-(p+q)$, so Eq.~\eqref{eq:n2right} reads
	\begin{equation}
		\begin{split}
			&K_{\rr,2}(q)\,S\bigl(-(p{+}q)\bigr)\,K_{\rr,1}(p)\,S\bigl(-(p{-}q)\bigr)\\
			&\qquad=S\bigl(-(p{-}q)\bigr)\,K_{\rr,1}(p)\,S\bigl(-(p{+}q)\bigr)\,K_{\rr,2}(q),
		\end{split}
		\label{eq:n2rightcentered}
	\end{equation}
	which is Eq.~\eqref{eq:REright}. With $\bar p=-p$ and $\bar q=-q$ it takes the corresponding form of the left reflection equation, as explained in Sec.~\ref{sec:factorization}.
	
	For the trivial right endpoint used in the impurity realizations, $K_\rr=\id$, and Eq.~\eqref{eq:n2right} collapses to
	\begin{equation}
		S(\beta+\Delta)\,S(\alpha)=S(\alpha)\,S(\beta+\Delta),
	\end{equation}
	which holds because the bulk matrices used here commute at different spectral arguments. Thus $K_\rr=\id$ is a solution of the right-endpoint relation for these particular $S$-matrices.

	\section{$O(3)$ bulk--impurity spectral normalization}
	\label{app:o3spectral}
	
	We can place the bulk and mixed matrices used in Sec.~\ref{sec:o3} on the same spectral scale directly. 
	
	Consider the rational matrix
	\begin{equation}
		R(w)=\id+\frac{\upsilon}{w}\PP-\frac{\upsilon}{w+\upsilon(N/2-1)}\Q .
		\label{eq:soNR}
	\end{equation}
	For $N=3$, its eigenvalues on $\mathcal P^{(2)},\mathcal P^{(1)},\mathcal P^{(0)}$ are $1+\upsilon/w$, $1-\upsilon/w$ and $(w-\upsilon)(w-\upsilon/2)/[w(w+\upsilon/2)]$. Normalizing by the first,
	\begin{equation}
		s_1=\frac{w-\upsilon}{w+\upsilon}, \qquad s_0=\frac{(w-\upsilon)(w-\upsilon/2)}{(w+\upsilon)(w+\upsilon/2)},
		\label{eq:O3scaleratios}
	\end{equation}
	and $w=\alpha u$ with $\upsilon/\alpha=-2i$ reproduces both. The antisymmetric channel has its pole at $u=2i$, whereas the corresponding $SU(N)$ pole sits at $u=i$. The same numerical drive rate $\kappa$ has different spectral units in the two models.
	
	For the vector--doublet matrix, let
	\begin{equation}
		L(w)=w\,\id_{a0}+\tfrac{\upsilon}{2}\,\vec\tau_a\cdot\vec\sigma_0 ,
		\label{eq:Lax}
	\end{equation}
	with vector particle space and doublet impurity space. It satisfies
	\begin{equation}
		R_{12}(w_1-w_2)L_1(w_1)L_2(w_2)=L_2(w_2)L_1(w_1)R_{12}(w_1-w_2)
		\label{eq:RLL}
	\end{equation}
	with $R$ of Eq.~\eqref{eq:soNR} at $N=3$, a standard intertwining relation. Setting
	\begin{equation}
		w=\alpha\bigl(u-\tfrac i2\bigr), \qquad \upsilon=-2i\alpha ,
		\label{eq:mixedshift}
	\end{equation}
	gives $L(w)=\alpha[(u-\tfrac i2)\id-iY_{a0}]$, exactly $\alpha$ times the numerator of Eq.~\eqref{eq:mixedO3}. The constant shift $-i\alpha/2$ is common to both factors in Eq.~\eqref{eq:RLL} and drops out of the difference $w_1-w_2$. Eqs.~\eqref{eq:O3scaleratios} and \eqref{eq:mixedshift} put the bulk and impurity matrices in the same spectral normalization~\cite{KattelO3SC2025}.
	
	\section{Scalar normalization and pure-exchange representatives}
	\label{app:normalization}
	
	Multiplying a solution of the Yang--Baxter equation by a scalar phase, or a solution of the reflection equation by a scalar phase, does not change the corresponding local factorization equation. The inverse Cayley transform is nonlinear and does change. This appendix derives the scalar choices used in the examples.
	
	Let a unitary eigenvalue be written as
	\begin{equation}
		\lambda_\mu=\frac{1-ia_\mu}{1+ia_\mu},\qquad x_\mu=a_\mu,
		\label{eq:normalization-channel}
	\end{equation}
	where $x_\mu$ is the Cayley eigenvalue in the convention of App.~\ref{app:transmission}. If the scalar phase is $\sigma=(1-ia_\sigma)/(1+ia_\sigma)$, multiplication of eigenvalues combines the corresponding Cayley parameters according to
	\begin{equation}
		a_\sigma\oplus a_\mu=\frac{a_\sigma+a_\mu}{1-a_\sigma a_\mu}.
		\label{eq:normalization-addition}
	\end{equation}
	If channel $\mu$ has dimension $n_\mu$, the trace condition is
	\begin{equation}
		\sum_\mu n_\mu\bigl(a_\sigma\oplus a_\mu\bigr)=0.
		\label{eq:tracequadratic}
	\end{equation}
	For a two-dimensional invariant algebra, this removes the identity component and leaves the unique traceless invariant exchange operator.
	
	For the rational $SU(N)$ bulk, the symmetric and antisymmetric channels have dimensions $n_\pm=N(N\pm1)/2$ and Cayley parameters $a_+=0$ and $a_-=-1/u$. Eq.~\eqref{eq:tracequadratic} becomes
	\begin{equation}
		(N+1)a_\sigma^2+2Nu\,a_\sigma-(N-1)=0.
		\label{eq:SUbulkquadratic}
	\end{equation}
	We choose
	\begin{equation}
		a_\sigma=a_N(u)=\frac{\sqrt{N^2u^2+N^2-1}-Nu}{N+1},
		\label{eq:aN}
	\end{equation}
	which is weakly coupled at one end of the real $u$ axis. The opposite root gives the other traceless representative. Define
	\begin{equation}
		\rho_N(u)=\frac{1-ia_N(u)}{1+ia_N(u)}.
		\label{eq:exchangephase}
	\end{equation}
	Then $\widetilde S_{12}(u)=\rho_N(u)S_{12}(u)$ has
	\begin{equation}
		\widetilde V_{12}(u)=g_N(u)t_1^At_2^A,\qquad g_N(u)=\frac{2N}{N-1}a_N(u).
		\label{eq:exchangeonly}
	\end{equation}
	This removes the scalar bulk term.
	
	The bulk scalar phase is also constrained by the global transport. Write a general rescaling as $\widetilde S(u)=\rho(u)S(u)$ and define $\omega(u)=\rho(u)\rho(-u)$. Repeating the two-particle flatness calculation with the scalar factors retained, the two orders of transport differ by
	\begin{equation}
		\frac{\omega(a)}{\omega(a+\Delta)},\qquad a=z_1-z_2.
		\label{eq:cocycle}
	\end{equation}
	Thus the rescaled bulk data preserve BqKZ flatness precisely when $\omega$ is $\Delta$-quasiconstant. The rational normalization has $\omega=1$. For the phase \eqref{eq:exchangephase}, $\omega$ is continuous and nonconstant and has finite limits as $u\to\pm\infty$, so it cannot be quasiconstant for nonzero $\Delta$. The corresponding pure-exchange bulk representative does not satisfy BqKZ flatness for $\Delta\neq0$.
	
	For the fundamental $SU(N)$ impurity, the two channels again have dimensions $N(N\pm1)/2$, now with $a_+=0$ and $a_-=-2u/D(u)$. The trace condition gives
	\begin{equation}
		(N+1)a_-a_\sigma^2-2Na_\sigma-(N-1)a_-=0.
		\label{eq:bdyquadratic}
	\end{equation}
	Choosing the branch that is weakly coupled for large $|u|$ gives Eqs.~\eqref{eq:BexchSUN} and \eqref{eq:gpartial}. $D(u)$ is even for every value of $d$. Away from $u=0$ the functions $g_\partial(u)$ and $a_\sigma(u)$ are odd, and for noncritical $d$ they extend through the origin whenever finite. Hence the scalar phase satisfies $\sigma(u)\sigma(-u)=1$ and preserves $K(u)K(-u)=\id$. At $d^2=-1$, the one-sided limits at $u=0$ have opposite signs. The corresponding scalar phase has no continuous extension through the origin that also preserves spectral reflection there, even though the original reflection matrix has a finite limit after canceling the common factor. At this critical value, the pure-exchange normalization is defined separately on the two sides of $u=0$ and does not provide a continuous family through the origin. In the BqKZ flatness relation each side contains the same reflection factors, so a scalar rescaling of $K$ cancels between the two sides wherever the normalized family is defined.
	
	For the $O(3)$ vector--doublet endpoint, the quartet and doublet channels have dimensions $4$ and $2$. Their Cayley parameters are $a_{3/2}=0$ and $a_{1/2}=-3u/E(u)$. Eq.~\eqref{eq:tracequadratic} becomes
	\begin{equation}
		2a_{1/2}a_\sigma^2-3a_\sigma-a_{1/2}=0.
		\label{eq:O3quadratic}
	\end{equation}
	The two roots can be written as
	\begin{equation}
		\gamma_\pm(u;d)=\frac{\pm\sqrt{E(u)^2+8u^2}-E(u)}{2u},
		\qquad \gamma_+\gamma_-=-2,
		\label{eq:gammaO3}
	\end{equation}
	where $\gamma=2a_\sigma$. $E(u)$ is even for every value of $d$. For $u\neq0$, both branches are odd. For noncritical $d$, this oddness extends through the origin whenever the branch is finite. At $d^2=-\tfrac94$, their one-sided limits at the origin have opposite signs. The scalar phase $\sigma=(1-i\gamma/2)/(1+i\gamma/2)$ has unequal one-sided limits and the exchange-normalized reflection matrix is discontinuous at $u=0$. Spectral reflection at $u=0$ would require $\sigma(0)^2=1$, which cannot be obtained by continuously extending either branch. The original rational reflection matrix remains finite there after canceling the common factor and taking the limit. The branch $\gamma_+$ behaves as $2/u$ for $u\to\pm\infty$, while $\gamma_-$ grows linearly. App.~\ref{app:cayleydomain} summarizes the finite-$u$ behavior.
	
	At a fixed spectral point $b>0$, write $J=J(b;d)$ and $E(b)=2b/c$. Then
	\begin{equation}
		d^2=b^2-\frac{2b}{c}-\frac94,
		\label{eq:dRG}
	\end{equation}
	and Eq.~\eqref{eq:JO3u} gives
	\begin{equation}
		c=\frac{2J}{1-2J^2}.
		\label{eq:cJ}
	\end{equation}
	Ref.~\cite{KattelO3SC2025} uses these parameter relations in the static parametrization.
	
	\section{Exceptional values of the reconstructed couplings}
	\label{app:cayleydomain}
	
	For the bulk representatives used in the driven models, the reconstructed local couplings become infinite at
	\begin{equation}
		\begin{split}
			V^{SU(N)}:&\quad u_B=0,\\
			\widetilde V^{O(3)}:&\quad u_B=0.
		\end{split}
		\label{eq:excluded-main}
	\end{equation}
	For the $SU(N)$ impurity, the rational coupling becomes infinite at $u_\ell=\pm\sqrt{1+d^2}$ when $1+d^2>0$. At $1+d^2=0$ these two points meet at $u_\ell=0$, while for $1+d^2<0$ the rational coupling remains finite on the real axis. The $O(3)$ impurity has the same structure with $1+d^2$ replaced by $d^2+\tfrac94$.
	
	These divergences come from the Cayley inversion, not from a singularity of the scattering or reflection matrix. At each such point one eigenvalue of the matching matrix reaches $U=-\id$, while $U$ itself remains finite and unitary. In a single eigenchannel,
	\begin{equation}
		U(x)=\frac{1-ix}{1+ix}=e^{2i\delta(x)}, \qquad \delta(x)=-\arctan x,
		\label{eq:cayleyphase}
	\end{equation}
	so $x\to\pm\infty$ corresponds to the finite matching value $U=-1$. Thus an infinite reconstructed coupling represents a finite matching condition rather than a singular scattering or reflection process.
	
	This distinction is important when the drive passes through one of these values. For $\kappa\neq0$, the affine coordinate $u_B(t)$ can reach such a point at a finite time $t_*$. Away from $t_*$, the matching matrix may be represented by the finite local coupling obtained from the Cayley inverse. At $t_*$, that Cayley coordinate is infinite, and the wavefunction is instead matched directly by the finite matrix $S$ or $K$. The factorized transport and the BqKZ relations are unchanged across an ordinary Cayley pole. When $\kappa=0$, the same issue arises only if the constant spectral argument is itself at one of these exceptional values.
	
	There is a separate issue for the scalar normalizations used to isolate pure exchange at the boundary. For the boundary impurities, the scalar normalization is compatible with the BqKZ transport, but at special parameter values the required scalar phase itself cannot be continued continuously through $u=0$.
	
	For the $SU(N)$ impurity, this occurs at $d^2=-1$. The pure-exchange coupling has finite one-sided limits, $g_\partial\to\pm\frac{4N}{\sqrt{N^2-1}}$, but the corresponding scalar phase has different limits as $u\to0^+$ and $u\to0^-$. Hence the pure-exchange reflection matrix cannot be continued continuously through $u=0$ while preserving $K(u)K(-u)=\id$. The original rational reflection matrix remains finite there.
	
	For the $O(3)$ impurity, the analogous critical value is $d^2=-\tfrac94$. At this point
	\begin{equation}
		\gamma_+\to\sqrt2\,\operatorname{sgn}u,\qquad \gamma_-\to-\sqrt2\,\operatorname{sgn}u,
	\end{equation}
	so the corresponding scalar phases again have different one-sided limits. Neither pure-exchange branch extends continuously through $u=0$ while preserving $K(u)K(-u)=\id$. The original rational $O(3)$ reflection matrix remains finite at the crossing.
	
	The two situations should be distinguished. At an ordinary Cayley pole, only the reconstructed local coupling becomes infinite, while the underlying scattering or reflection matrix remains finite. The factorized description continues through the matching matrix. At the critical pure-exchange points $d^2=-1$ and $d^2=-\tfrac94$, the exchange-normalized reflection matrix itself has different one-sided limits at $u=0$. Hence, that pure-exchange family does not define a continuous continuation through the crossing. The original rational reflection matrix remains regular at the same point, but it is a different scalar normalization rather than a continuation of the pure-exchange family. App.~\ref{app:normalization} gives the corresponding scalar branches.
	
	\section{Verification of the boundary qKZ equations}
	\label{app:bqkzproof}
	
	We verify the boundary qKZ equation for $j=n$ and $M=1$ on a generic lattice, and then indicate the extensions. The bulk factors are $S\equiv S_\eta$, the impurity is the site $0$ with $z_0=d$, and $K_\ell=K_\rr=\id$ after this replacement. All sums are taken at $\operatorname{Im}\eta>0$, where they converge absolutely.
	
	\paragraph*{Double row on the reference state.}
	Each bulk factor can be written as
	\begin{equation}
		S_{ar}(x)=\frac{x}{x-i\eta}\,\id-\frac{i\eta}{x-i\eta}\,\PP_{ar}.
	\end{equation}
	It acts as the identity when the auxiliary spin and the spin at site $r$ are equal. On the two mixed states it keeps the spins with amplitude $x/(x-i\eta)$ and exchanges them with amplitude $-i\eta/(x-i\eta)$. In $\mathbb T_a(-\theta)$ the auxiliary spin meets the sites in the order $n,\ldots,0$, and in $\mathbb T_a(\theta)^{-1}$ in the order $0,\ldots,n$. If it stays up, every factor acts as the identity, which gives $\mathsf A(\theta)|{\Uparrow}\rangle=|{\Uparrow}\rangle$ and $\mathsf C(\theta)|{\Uparrow}\rangle=0$. The entry $\mathsf D$ has two contributions. If the auxiliary spin stays down on both rows, the diagonal amplitudes multiply to
	\begin{equation}
		g_r(\theta)=\frac{(\theta-z_r)(\theta+z_r)}{(\theta-z_r+i\eta)(\theta+z_r+i\eta)}
	\end{equation}
	for each site, and their product is $\mathfrak g(\theta)$. Otherwise the auxiliary spin flips site $r$ on the first row and flips it back on the second. This path contributes
	\begin{equation}
		\begin{split}
			&\frac{(i\eta)^2}{(\theta-z_r+i\eta)(\theta+z_r+i\eta)}\prod_{l>r}g_l(\theta)\\
			&\qquad=\frac{i\eta}{2\theta+i\eta}\bigl[1-g_r(\theta)\bigr]\prod_{l>r}g_l(\theta),
		\end{split}
	\end{equation}
	where $(\theta+i\eta)^2-\theta^2=i\eta(2\theta+i\eta)$ was used. The sum over $r$ telescopes to $1-\mathfrak g(\theta)$, which gives Eq.~\eqref{eq:bdyvacuumeig}. The constant part is removed by
	\begin{equation}
		\widetilde{\mathsf D}(\theta)=\mathsf D(\theta)-\frac{i\eta}{2\theta+i\eta}\mathsf A(\theta),\qquad
		\widetilde{\mathsf D}(\theta)|{\Uparrow}\rangle=\frac{2\theta\,\mathfrak g(\theta)}{2\theta+i\eta}|{\Uparrow}\rangle ,
	\end{equation}
	so that
	\begin{equation}
		\mathsf t(\theta)=\frac{2(\theta+i\eta)}{2\theta+i\eta}\mathsf A(\theta)+\widetilde{\mathsf D}(\theta).
		\label{eq:tAD}
	\end{equation}
	Writing out $g_r(\theta-i\eta/2)$ also shows
	\begin{equation}
		\mathfrak g\bigl(\theta-\tfrac{i\eta}{2}\bigr)=\frac{\Upsilon(-\theta)}{\Upsilon(\theta)} .
		\label{eq:gUpsilon}
	\end{equation}
	
	\paragraph*{Reflection algebra and exchange relations.}
	The Yang--Baxter equation, applied site by site, gives $S_{12}(v-y)\mathbb T_1(v)\mathbb T_2(y)=\mathbb T_2(y)\mathbb T_1(v)S_{12}(v-y)$ for two auxiliary spaces $1$ and $2$. Multiplying by inverses and using $[S_{12}(x),S_{12}(x')]=0$, one obtains the reflection algebra
	\begin{equation}
		\begin{split}
			&S_{12}(y-v)\,\mathbb U_1(v)\,S_{12}(-v-y)\,\mathbb U_2(y)\\
			&\qquad=\mathbb U_2(y)\,S_{12}(-v-y)\,\mathbb U_1(v)\,S_{12}(y-v).
		\end{split}
		\label{eq:reflalg}
	\end{equation}
	Its matrix elements give $[\mathsf B(v),\mathsf B(y)]=0$ and, in terms of $\widetilde{\mathsf D}$, the exchange relations~\cite{Sklyanin1988}
	\begin{equation}
		\begin{split}
			\mathsf A(v)\mathsf B(y)={}&\frac{(v-y-i\eta)(v+y)}{(v-y)(v+y+i\eta)}\mathsf B(y)\mathsf A(v)\\
			&+\frac{2i\eta\,y}{(v-y)(2y+i\eta)}\mathsf B(v)\mathsf A(y)\\
			&-\frac{i\eta}{v+y+i\eta}\mathsf B(v)\widetilde{\mathsf D}(y),
		\end{split}
		\label{eq:ABex}
	\end{equation}
	\begin{equation}
		\begin{split}
			\widetilde{\mathsf D}(v)\mathsf B(y)={}&\frac{(v-y+i\eta)(v+y+2i\eta)}{(v-y)(v+y+i\eta)}\mathsf B(y)\widetilde{\mathsf D}(v)\\
			&-\frac{2i\eta(v+i\eta)}{(v-y)(2v+i\eta)}\mathsf B(v)\widetilde{\mathsf D}(y)\\
			&+\frac{4i\eta\,y(v+i\eta)}{(v+y+i\eta)(2v+i\eta)(2y+i\eta)}\mathsf B(v)\mathsf A(y).
		\end{split}
		\label{eq:DBex}
	\end{equation}
	In each relation the first term keeps the argument $y$ of the creation operator, while in the other terms the creation operator has taken the argument $v$.
	
	\paragraph*{Off-shell action.}
	Acting with Eq.~\eqref{eq:tAD} on $\mathsf B(y)|{\Uparrow}\rangle$ and using Eqs.~\eqref{eq:ABex} and \eqref{eq:DBex}, together with the vacuum eigenvalues, the terms with $\mathsf B(y)|{\Uparrow}\rangle$ give the wanted coefficient. The terms with $\mathsf B(v)|{\Uparrow}\rangle$ combine, through $1/(v-y)+1/(v+y+i\eta)=(2v+i\eta)/[(v-y)(v+y+i\eta)]$, into the unwanted coefficient
	\begin{equation}
		\frac{2i\eta(v+i\eta)}{(v-y)(v+y+i\eta)}\left[\frac{2y}{2y+i\eta}-\frac{2y\,\mathfrak g(y)}{2y+i\eta}\right].
	\end{equation}
	For $y=\theta-i\eta/2$, Eq.~\eqref{eq:gUpsilon} and the normalization~\eqref{eq:Bbar} turn this into Eq.~\eqref{eq:bdyoffshellaction}, with the wanted coefficient
	\begin{equation}
		\begin{split}
			\mathscr L(v;\theta)={}&\frac{2(v+i\eta)}{2v+i\eta}\frac{(v-\theta-i\eta/2)(v+\theta-i\eta/2)}{(v-\theta+i\eta/2)(v+\theta+i\eta/2)}\\
			&+\frac{2v\,\mathfrak g(v)}{2v+i\eta}\frac{(v-\theta+3i\eta/2)(v+\theta+3i\eta/2)}{(v-\theta+i\eta/2)(v+\theta+i\eta/2)} .
		\end{split}
		\label{eq:wantedL}
	\end{equation}
	Both coefficients are even in $\theta$, as required by the oddness of $\bar{\mathsf B}(\theta)$. For $M$ roots, the wanted coefficient is the same expression with each $\theta$-dependent fraction replaced by its product over the roots, and the unwanted coefficient of root $m$ is $2i\eta(v+i\eta)[\mathcal X_m(\theta_m)-\mathcal X_m(-\theta_m)]/[(v-\theta_m+i\eta/2)(v+\theta_m+i\eta/2)]$, multiplying $\mathsf B(v)\prod_{k\ne m}\bar{\mathsf B}(\theta_k)|{\Uparrow}\rangle$.
	
	\paragraph*{The last round trip as a trace.}
	For $j=n$, with the impurity as site $0$,
	\begin{equation}
		\begin{split}
			\mathcal T_n(\bm z)={}&S_{n,n-1}(z_n+z_{n-1})\cdots S_{n0}(z_n+z_0)\\
			&\times S_{n0}(z_n-z_0)\cdots S_{n,n-1}(z_n-z_{n-1}).
		\end{split}
		\label{eq:Tnexplicit}
	\end{equation}
	At $\theta=-z_n$ the double row reads $\mathbb U_a(-z_n)=S_{an}(2z_n)X_aS_{an}(0)$, where $X_a$ is Eq.~\eqref{eq:Tnexplicit} with the label $n$ replaced by $a$. The last factor is $S_{an}(0)=\PP_{an}$, and $X_a\PP_{an}=\PP_{an}X_n=\PP_{an}\mathcal T_n$. Since $\mathcal T_n$ does not act on the auxiliary space,
	\begin{equation}
		\mathsf t(-z_n)=\Tr_a\bigl[S_{an}(2z_n)\PP_{an}\bigr]\mathcal T_n(\bm z)=\frac{2(z_n-i\eta)}{2z_n-i\eta}\,\mathcal T_n(\bm z),
	\end{equation}
	because $S_{an}(x)\PP_{an}=(ix\PP_{an}+\eta)/(ix+\eta)$, $\Tr_a\PP_{an}=\id$, and $\Tr_a\id=2$. This is Eq.~\eqref{eq:Tntransfer}. It holds for every $\Delta$ and uses only $S(0)=\PP$ and braiding unitarity. This is the only point at which the transport and the auxiliary double row are compared. The exchange relations~\eqref{eq:ABex} and~\eqref{eq:DBex} are properties of $\mathbb U_a$ alone; they do not refer to the transport and need no dual reflection equation, which only guarantees that traces at different spectral parameters commute. The proof below uses the algebraic Bethe ansatz for exactly one operator, $\mathsf t(-z_n)$, which by Eq.~\eqref{eq:Tntransfer} is the transport of the last electron.
	
	The identity requires an empty return leg and a trivial far reflection. With a nontrivial $K_\rr$, a trace closed with a matrix $K^+$ produces on site $n$ the operator $\Tr_a[K^+_aS_{an}(2z_n)\PP_{an}]\propto 2iz_nK^+_n+\eta\Tr K^+$, multiplying the same product of bulk factors, while the transport has $K_\rr(-z_n-\Delta/2)$ there. Take $K_\rr$ of the form~\eqref{eq:Kcnumbernorm} with $M_\rr\ne\pm\id$ and $\xi_\rr\ne0$; for $SU(2)$, $M_\rr^2=\id$ then implies $\Tr M_\rr=0$. The traceless part of $2iz_nK^++\eta\Tr K^+$ is $2iz_n$ times that of $K^+$, so the two can be proportional only if $K^+\propto\xi_\rr\id+iwM_\rr$ for some $w$. Then
	\begin{equation}
		2iz_nK^++\eta\Tr K^+\propto\xi_\rr\id-\frac{z_nw}{\eta+iz_n}\,M_\rr ,
	\end{equation}
	and comparison with $K_\rr(-z_n-\Delta/2)\propto\xi_\rr\id-i(z_n+\Delta/2)M_\rr$ fixes $w$. With $\theta=-z_n$,
	\begin{equation}
		K^+(\theta)=K_\rr\bigl((\theta+i\eta)(1-\Delta/2\theta)\bigr).
		\label{eq:Kplusdelta}
	\end{equation}
	At $\Delta=0$ this is $K^+(\theta)=K_\rr(\theta+i\eta)$, the crossing dual of Eq.~\eqref{eq:staticdual}. The corresponding traces commute, and at $\Delta=0$ the trace reproduces every $\mathcal T_j^{(0)}$, not only $j=n$. At $\Delta\ne0$ the matrix~\eqref{eq:Kplusdelta} depends on $\Delta$, and the traces it closes do not commute, so it is not a solution of the dual reflection equation. The excluded case $\xi_\rr=0$ is the constant matrix $K_\rr=M_\rr$, for which $K^+=M_\rr$ works at every $\Delta$. The round trip with a nontrivial far reflection is a genuinely different operator from Sklyanin's transfer matrix at finite $\Delta$. This is why the identity is used here only for $K_\rr=\id$, where the far end of both objects is the identity.
	
	\paragraph*{Action of the last round trip.}
	At $v=-z_n$, $\mathfrak g(-z_n)=0$ because $g_n$ contains $\theta+z_n$, so the second line of Eq.~\eqref{eq:wantedL} drops out. The prefactor in Eq.~\eqref{eq:Tntransfer} cancels $2(v+i\eta)/(2v+i\eta)$ at $v=-z_n$. Acting with $\mathbb U_a(-z_n)$ on $|\downarrow\rangle_a|{\Uparrow}\rangle$, the permutation gives $|\uparrow\rangle_a\sigma_n^-|{\Uparrow}\rangle$, $X_a$ then acts as the identity, and $S_{an}(2z_n)$ keeps the spins with amplitude $2z_n/(2z_n-i\eta)$. Hence $\mathsf B(-z_n)|{\Uparrow}\rangle=[2z_n/(2z_n-i\eta)]\sigma_n^-|{\Uparrow}\rangle$, the analog of $\mathsf B(\xi_1)|{\Uparrow}\rangle=a(\xi_1)\sigma_1^-|{\Uparrow}\rangle$ in the periodic problem. These three facts give Eqs.~\eqref{eq:Tnaction} and~\eqref{eq:LnUn}, and 
	\begin{equation}
		\mathcal T_nF=\sum_{\theta\in\Lambda}\Phi_1(\bm z;\theta)\Bigl[\mathscr L_n(\theta)\bar{\mathsf B}(\theta)|{\Uparrow}\rangle+\mathscr U_n(\theta)\sigma_n^-|{\Uparrow}\rangle\Bigr].
	\end{equation}
	
	\paragraph*{Shifted amplitude.}
	Under $z_n\to z_n+\Delta$, Eq.~\eqref{eq:bdyWshift} gives
	\begin{equation}
		\begin{split}
			\frac{\Phi_1(\bm z+\Delta e_n;\theta)}{\Phi_1(\bm z;\theta)}={}&\frac{\theta+z_n+\Delta+i\eta/2}{\theta+z_n+\Delta-i\eta/2}\\
			&\times\frac{\theta-z_n-\Delta-i\eta/2}{\theta-z_n-\Delta+i\eta/2}.
		\end{split}
		\label{eq:weightshiftn}
	\end{equation}
	For $r<n$, $z_n$ enters $e_r(\theta)$ of Eq.~\eqref{eq:ek} through $(\theta-z_n+i\eta/2)/(\theta-z_n-i\eta/2)$ in $\Upsilon$ and $(\theta+z_n-i\eta/2)/(\theta+z_n+i\eta/2)$ in the product. After the shift, the $\Delta$-dependent parts of these two factors cancel the two factors of Eq.~\eqref{eq:weightshiftn}, and what remains is
	\begin{equation}
		\frac{\theta-z_n-i\eta/2}{\theta-z_n+i\eta/2}\,\frac{\theta+z_n+i\eta/2}{\theta+z_n-i\eta/2}=\mathscr L_n(\theta).
	\end{equation}
	Since $\mathscr L_n$ is even, the same holds for $e_r(-\theta)$. The components $r<n$ of $F(\bm z+\Delta e_n)$ coincide with those of $\mathcal T_nF$ term by term.
	
	For $r=n$, let $\Upsilon_{<n}(\theta)=\prod_{r<n}(\theta-z_r+i\eta/2)/(\theta-z_r-i\eta/2)$, so that $e_n(\theta)=i\eta\Upsilon_{<n}(\theta)(\theta-z_n+i\eta/2)/[(\theta-z_n-i\eta/2)(\theta+z_n+i\eta/2)]$. After the shift, the factor $(\theta-z_n-\Delta+i\eta/2)/(\theta-z_n-\Delta-i\eta/2)$ cancels the second factor of Eq.~\eqref{eq:weightshiftn}, and $1/(\theta+z_n+\Delta+i\eta/2)$ combines with the first. The $n$th component of $F(\bm z+\Delta e_n)$ is 
	\begin{equation}
		\begin{split}
			i\eta\sum_\theta\Phi_1(\bm z;\theta)\biggl[&\frac{\Upsilon_{<n}(\theta)}{\theta+z_n+\Delta-i\eta/2}\\
			&-\frac{\Upsilon_{<n}(-\theta)}{-\theta+z_n+\Delta-i\eta/2}\biggr].
		\end{split}
	\end{equation}
	On the right side, $\mathscr L_n(\theta)e_n(\theta)=i\eta\Upsilon_{<n}(\theta)/(\theta+z_n-i\eta/2)$. Adding $\mathscr U_n$ written with $\Upsilon_{<n}$, the coefficient of $i\eta\Upsilon_{<n}(\theta)$ becomes $1/(\theta-z_n-i\eta/2)$ and that of $i\eta\Upsilon_{<n}(-\theta)$ becomes $1/(\theta+z_n+i\eta/2)$:
	\begin{equation}
		\begin{split}
			(\mathcal T_nF)_n=i\eta\sum_\theta\Phi_1(\bm z;\theta)\biggl[&\frac{\Upsilon_{<n}(\theta)}{\theta-z_n-i\eta/2}\\
			&+\frac{\Upsilon_{<n}(-\theta)}{\theta+z_n+i\eta/2}\biggr].
		\end{split}
	\end{equation}
	Subtracting the two and returning from $\Upsilon_{<n}$ to $\Upsilon$ gives the two sums of Eq.~\eqref{eq:bdyunwanted} with the prefactor $i\eta(2z_n+\Delta)$. By Eq.~\eqref{eq:bdykeyid}, the second sum equals the first after the shift $\theta\to\theta-\Delta$ of the summation variable. Equivalently, the difference of the two summands is $\mathscr H(\theta)-\mathscr H(\theta+\Delta)$ with
	\begin{equation}
		\mathscr H(\theta)=i\eta(2z_n+\Delta)\frac{\Phi_1(\bm z;\theta)\,\Upsilon(-\theta)}{(\theta+z_n-i\eta/2)(\theta-z_n-\Delta+i\eta/2)} ,
	\end{equation}
	so the symmetric partial sum telescopes to $\mathscr H(\theta^0-k_{\rm c}\Delta)-\mathscr H(\theta^0+(k_{\rm c}+1)\Delta)$. Since $|\Phi_1|$ stays bounded and $\mathscr H=O(\theta^{-2})$, both end terms vanish. A generic lattice is invariant under the shift, and both sums converge absolutely for $\operatorname{Im}\eta>0$. This proves Eq.~\eqref{eq:BqKZ} for $j=n$ in the region of absolute convergence and hence, by continuation, at the physical point.
	
	\paragraph*{Exchange and the remaining equations.}
	Let $s_b$ exchange the electron labels $z_b$ and $z_{b+1}$ and let $\check R_b(\bm z)=\PP_{b,b+1}S_{b,b+1}(z_{b+1}-z_b)$. The Yang--Baxter equation gives $\check R_b(\bm z)\,\mathbb T_a(v;\bm z)=\mathbb T_a(v;s_b\bm z)\,\check R_b(\bm z)$, and hence the same relation for $\mathbb U_a$ and for $\mathsf B$. Since $\Upsilon$ and $\Phi$ are symmetric in the labels and $\check R_b|{\Uparrow}\rangle=|{\Uparrow}\rangle$, Eq.~\eqref{eq:bdybosonic} follows, $F(s_b\bm z)=\check R_b(\bm z)F(\bm z)$. Relabeling the electrons $b$ and $b+1$ in the round trip and canceling their crossing at the two ends by braiding unitarity gives
	\begin{equation}
		\mathcal T_b(\bm z)=\check R_b(s_b\bm z+\Delta e_{b+1})\,\mathcal T_{b+1}(s_b\bm z)\,\check R_b(\bm z).
	\end{equation}
	Since $\bm z+\Delta e_b=s_b(s_b\bm z+\Delta e_{b+1})$,
	\begin{equation}
		\begin{split}
			F(\bm z+\Delta e_b)&=\check R_b(s_b\bm z+\Delta e_{b+1})F(s_b\bm z+\Delta e_{b+1})\\
			&=\check R_b(s_b\bm z+\Delta e_{b+1})\mathcal T_{b+1}(s_b\bm z)\check R_b(\bm z)F(\bm z)\\
			&=\mathcal T_b(\bm z)F(\bm z).
		\end{split}
	\end{equation}
	Starting from $b+1=n$, this gives the equations for $j=n-1,\ldots,1$.
	
	\paragraph*{Several roots and highest weight.}
	For $M\ge2$, the action of $\mathsf t(-z_n)$ on $\Omega$ produces, for each $m$, an unwanted term proportional to $\mathcal X_m(\theta_m)-\mathcal X_m(-\theta_m)$. The shift relation~\eqref{eq:bdyGshift} gives, for both signs and both orderings of $m$ and $k$,
	\begin{equation}
		\begin{split}
			&\prod_{\upsilon=\pm}\frac{\theta_m+\upsilon\theta_k-i\eta}{\theta_m+\upsilon\theta_k}\,\frac{\theta_m+\upsilon\theta_k+\Delta}{\theta_m+\upsilon\theta_k+\Delta+i\eta}\\
			&\qquad=\frac{\mathfrak r(\theta_m,\theta_k)}{\mathfrak r(-\theta_m-\Delta,\theta_k)},
		\end{split}
	\end{equation}
	which, combined with the one-root identity, is Eq.~\eqref{eq:bdykeyidM}. This identity is what is needed for the weight $\Phi$ to convert each unwanted term into a total difference in $\theta_m$, as in the case $M=1$ and as in the diagonal-boundary construction of Ref.~\cite{ReshetikhinStokmanVlaar2015}. The bookkeeping of the components is longer, because the shifted label $z_n$ also enters the other creation operators. The highest-weight property is $S^+_{\rm tot}F_\Lambda=0$. As $\Delta\to0$, $\Phi(\bm z,\bm\theta+\Delta e_m)/\Phi(\bm z,\bm\theta)\to\mathcal X_m(\theta_m)/\mathcal X_m(-\theta_m)$, so stationarity of the weight, used in Sec.~\ref{sec:smallstep}, is the vanishing of the unwanted coefficients, which are the ordinary Bethe equations.
	
	\bibliography{references}

@article{ZZ1979,
 author = {Zamolodchikov, Alexander B. and Zamolodchikov, Alexey B.},
 title = {Factorized {$S$}-matrices in two dimensions as the exact solutions
 of certain relativistic quantum field theory models},
 journal = {Ann. Phys. (N.Y.)},
 volume = {120},
 number = {2},
 pages = {253--291},
 year = {1979},
 doi = {10.1016/0003-4916(79)90391-9}
}

@misc{Faddeev1996,
 author = {Faddeev, Ludwig D.},
 title = {How algebraic {B}ethe ansatz works for integrable model},
 year = {1996},
 eprint = {hep-th/9605187},
 archivePrefix= {arXiv},
 primaryClass = {hep-th}
}

@article{Cherednik1984,
 author = {Cherednik, Ivan V.},
 title = {Factorizing particles on a half-line and root systems},
 journal = {Theor. Math. Phys.},
 volume = {61},
 number = {1},
 pages = {977--983},
 year = {1984},
 doi = {10.1007/BF01038545}
}

@article{Sklyanin1988,
 author = {Sklyanin, Evgeny K.},
 title = {Boundary conditions for integrable quantum systems},
 journal = {J. Phys. A: Math. Gen.},
 volume = {21},
 number = {10},
 pages = {2375--2389},
 year = {1988},
 doi = {10.1088/0305-4470/21/10/015}
}

@article{GhoshalZamolodchikov1994,
 author = {Ghoshal, Subir and Zamolodchikov, Alexander B.},
 title = {Boundary {$S$}-matrix and boundary state in two-dimensional
 integrable quantum field theory},
 journal = {Int. J. Mod. Phys. A},
 volume = {9},
 number = {21},
 pages = {3841--3885},
 year = {1994},
 doi = {10.1142/S0217751X94001552}
}

@article{FrenkelReshetikhin1992,
 author = {Frenkel, Igor B. and Reshetikhin, Nicolai Yu.},
 title = {Quantum affine algebras and holonomic difference equations},
 journal = {Commun. Math. Phys.},
 volume = {146},
 number = {1},
 pages = {1--60},
 year = {1992},
 doi = {10.1007/BF02099206}
}

@article{Cherednik1992,
 author = {Cherednik, Ivan V.},
 title = {Quantum {K}nizhnik--{Z}amolodchikov equations and affine root
 systems},
 journal = {Commun. Math. Phys.},
 volume = {150},
 number = {1},
 pages = {109--136},
 year = {1992},
 doi = {10.1007/BF02096568}
}

@misc{KattelGeneral2026,
 author = {Kattel, Pradip},
 title = {General construction of time-dependent integrable chiral
 field theories},
 year = {2026},
 eprint = {2608.23770},
 archivePrefix= {arXiv},
 primaryClass = {cond-mat.str-el}
}

@misc{PasnooriBoundaryqKZ2026,
 author = {Pasnoori, Parameshwar R.},
 title = {Boundary quantum {K}nizhnik--{Z}amolodchikov equations and
 integrability of quantum field theories with time-dependent
 bulk and boundary coupling strengths},
 year = {2026},
 eprint = {2608.31080},
 archivePrefix= {arXiv},
 primaryClass = {cond-mat.str-el}
}

@article{PasnooriKondo2025,
 author = {Pasnoori, Parameshwar R.},
 title = {Integrability of the {K}ondo model with time-dependent
 interaction strength},
 journal = {Phys. Rev. B},
 volume = {112},
 pages = {L060409},
 year = {2025},
 eprint = {2505.20125},
 archivePrefix = {arXiv}
}

@article{KattelO3SC2025,
 author = {Kattel, Pradip and Zhakenov, Abay and Andrei, Natan},
 title = {Kondo overscreening in the presence of superconductivity},
 journal = {Phys. Rev. B},
 volume = {112},
 pages = {085103},
 year = {2025},
 doi = {10.1103/rg1x-bztv},
 eprint = {2412.01924},
 archivePrefix = {arXiv}
}

@article{KattelXXZ2025,
 author = {Kattel, Pradip and Pasnoori, Parameshwar R. and Pixley, J. H. and
 Andrei, Natan},
 title = {Edge modes and boundary impurities in the anisotropic
 {H}eisenberg spin chain},
 journal = {Phys. Rev. B},
 volume = {111},
 pages = {174430},
 year = {2025},
 eprint = {2412.00194},
 archivePrefix = {arXiv}
}

@article{PasnooriRylandsAndrei2020,
 author = {Pasnoori, Parameshwar R. and Rylands, Colin and Andrei, Natan},
 title = {Kondo impurity at the edge of a superconducting wire},
 journal = {Phys. Rev. Research},
 volume = {2},
 pages = {013006},
 year = {2020},
 doi = {10.1103/PhysRevResearch.2.013006}
}

@article{PasnooriYSR2022,
 author = {Pasnoori, Parameshwar R. and Andrei, Natan and Rylands, Colin and
 Azaria, Patrick},
 title = {Rise and fall of {Y}u--{S}hiba--{R}usinov bound states in
 charge-conserving {$s$}-wave one-dimensional superconductors},
 journal = {Phys. Rev. B},
 volume = {105},
 pages = {174517},
 year = {2022},
 doi = {10.1103/PhysRevB.105.174517}
}

@article{AndreiDestri1984,
 author = {Andrei, Natan and Destri, Claudio},
 title = {Dynamical symmetry breaking and fractionization in a new
 integrable model},
 journal = {Nucl. Phys. B},
 volume = {231},
 number = {3},
 pages = {445--480},
 year = {1984},
 doi = {10.1016/0550-3213(84)90515-2}
}

@article{AndreiFuruyaLowenstein1983,
 author = {Andrei, Natan and Furuya, Katsuhiro and Lowenstein, John H.},
 title = {Solution of the {K}ondo problem},
 journal = {Rev. Mod. Phys.},
 volume = {55},
 number = {2},
 pages = {331--402},
 year = {1983},
 doi = {10.1103/RevModPhys.55.331}
}

@article{MezincescuNepomechie1992,
 author = {Mezincescu, Luca and Nepomechie, Rafael I.},
 title = {Fusion procedure for open chains},
 journal = {J. Phys. A: Math. Gen.},
 volume = {25},
 number = {9},
 pages = {2533--2543},
 year = {1992},
 doi = {10.1088/0305-4470/25/9/024}
}

@book{WangYangCaoShi2015,
 author = {Wang, Yupeng and Yang, Wen-Li and Cao, Junpeng and Shi, Kangjie},
 title = {Off-Diagonal {B}ethe Ansatz for Exactly Solvable Models},
 publisher = {Springer},
 address = {Berlin, Heidelberg},
 year = {2015},
 doi = {10.1007/978-3-662-46756-5}
}

@article{FringKoberleToda1994,
 author = {Fring, Andreas and K{\"o}berle, Roland},
 title = {Affine Toda Field Theory in the Presence of Reflecting Boundaries},
 journal = {Nuclear Physics B},
 volume = {419},
 pages = {647--664},
 year = {1994},
 eprint = {hep-th/9309142},
 archivePrefix = {arXiv},
 doi = {10.1016/0550-3213(94)90349-2}
}

@article{GhoshalON1994,
 author = {Ghoshal, Subir},
 title = {Boundary $S$ Matrix of the $O(N)$ Symmetric Nonlinear Sigma Model},
 journal = {Physics Letters B},
 volume = {334},
 pages = {363--368},
 year = {1994},
 eprint = {hep-th/9401008},
 archivePrefix = {arXiv},
 doi = {10.1016/0370-2693(94)90701-3}
}

@article{DeMartinoMoriconi1999,
 author = {De Martino, A. and Moriconi, M.},
 title = {Boundary $S$ Matrix for the Gross--Neveu Model},
 journal = {Physics Letters B},
 volume = {451},
 pages = {354--364},
 year = {1999},
 eprint = {hep-th/9812009},
 archivePrefix = {arXiv},
 doi = {10.1016/S0370-2693(99)00240-3}
}

@article{LeClairMussardoSaleurSkorik1995,
 author = {LeClair, A. and Mussardo, G. and Saleur, H. and Skorik, S.},
 title = {Boundary Energy and Boundary States in Integrable Quantum Field Theories},
 journal = {Nuclear Physics B},
 volume = {453},
 pages = {581--618},
 year = {1995},
 eprint = {hep-th/9503227},
 archivePrefix = {arXiv},
 doi = {10.1016/0550-3213(95)00435-U}
}

@article{DoreyPocklingtonTateoWatts1998,
 author = {Dorey, Patrick and Pocklington, Andrew and Tateo, Roberto and Watts, Gerard},
 title = {{TBA} and {TCSA} with Boundaries and Excited States},
 journal = {Nuclear Physics B},
 volume = {525},
 pages = {641--663},
 year = {1998},
 eprint = {hep-th/9712197},
 archivePrefix = {arXiv},
 doi = {10.1016/S0550-3213(98)00339-3}
}

@article{DoreyRunkelTateoWatts2000,
 author = {Dorey, Patrick and Runkel, Ingo and Tateo, Roberto and Watts, Gerard M. T.},
 title = {$g$-Function Flow in Perturbed Boundary Conformal Field Theories},
 journal = {Nuclear Physics B},
 volume = {578},
 pages = {85--122},
 year = {2000},
 eprint = {hep-th/9909216},
 archivePrefix = {arXiv},
 doi = {10.1016/S0550-3213(99)00772-5}
}

@article{kattel2026multichannel,
 title={Multichannel Kondo effect in one-dimensional superconducting leads},
 author={Kattel, Pradip and Zhakenov, Abay and Andrei, Natan},
 journal={Physical Review B},
 volume={113},
 number={16},
 pages={165130},
 year={2026},
 publisher={APS}
}

@article{rylands2020exact,
 title={Exact boundary modes in an interacting quantum wire},
 author={Rylands, Colin},
 journal={Physical Review B},
 volume={101},
 number={8},
 pages={085133},
 year={2020},
 publisher={APS}
}

@article{SinitsynYuzbashyan2018,
  title = {Integrable Time-Dependent Quantum Hamiltonians},
  author = {Sinitsyn, Nikolai A. and Yuzbashyan, Emil A. and Chernyak, Vladimir Y. and Patra, Aniket and Sun, Chen},
  journal = {Physical Review Letters},
  volume = {120},
  number = {19},
  pages = {190402},
  year = {2018},
  publisher = {American Physical Society},
  doi = {10.1103/PhysRevLett.120.190402}
}

@article{Yuzbashyan2018,
  title = {Integrable time-dependent Hamiltonians, solvable Landau--Zener models and Gaudin magnets},
  author = {Yuzbashyan, Emil A.},
  journal = {Annals of Physics},
  volume = {392},
  pages = {323--339},
  year = {2018},
  publisher = {Elsevier},
  doi = {10.1016/j.aop.2018.01.017}
}

@article{ReshetikhinStokmanVlaar2015,
  author = {Reshetikhin, Nicolai and Stokman, Jasper and Vlaar, Bart},
  title = {Boundary Quantum Knizhnik--Zamolodchikov Equations and Bethe Vectors},
  journal = {Communications in Mathematical Physics},
  volume = {336},
  number = {2},
  pages = {953--986},
  year = {2015},
  doi = {10.1007/s00220-014-2227-2},
  eprint = {1305.1113},
  archivePrefix = {arXiv},
  primaryClass = {math.QA}
}

@article{ReshetikhinStokmanVlaar2018,
  author = {Reshetikhin, Nicolai and Stokman, Jasper and Vlaar, Bart},
  title = {Integral Solutions to Boundary Quantum Knizhnik--Zamolodchikov Equations},
  journal = {Advances in Mathematics},
  volume = {323},
  pages = {486--528},
  year = {2018},
  doi = {10.1016/j.aim.2017.10.041},
  eprint = {1602.08457},
  archivePrefix = {arXiv},
  primaryClass = {math.QA}
}

@article{andrei1979diagonalization,
  title={Diagonalization of the chiral-invariant Gross-Neveu Hamiltonian},
  author={Andrei, N},
  journal={Physical review letters},
  volume={43},
  number={23},
  pages={1698},
  year={1979},
  publisher={APS}
}

@misc{MukhinVarchenko1997,
  author = {Mukhin, E. and Varchenko, A.},
  title  = {The quantized {K}nizhnik-{Z}amolodchikov equation in tensor products of irreducible $sl_2$-modules},
  note   = {arXiv:q-alg/9709026}, year = {1997}}

@article{ReshetikhinStokmanVlaarFusion2016,
  author  = {Reshetikhin, N. and Stokman, J. and Vlaar, B.},
  title   = {Boundary quantum {K}nizhnik-{Z}amolodchikov equations and fusion},
  journal = {Ann. Henri Poincar\'e}, volume = {17}, pages = {137}, year = {2016},
  doi     = {10.1007/s00023-014-0395-4}, note = {arXiv:1404.5492}}

@article{Reshetikhin1992Jackson,
  author  = {Reshetikhin, Nicolai},
  title   = {Jackson-type integrals, Bethe vectors, and solutions to a difference analog of the Knizhnik--Zamolodchikov system},
  journal = {Letters in Mathematical Physics},
  volume  = {26},
  pages   = {153--165},
  year    = {1992},
  month   = nov,
  doi     = {10.1007/BF00420749}
}

@article{kattel2026integrable,
  title={Integrable Chiral Quantum Field Theories with Spacetime-Dependent Interactions},
  author={Kattel, Pradip},
  journal={arXiv preprint arXiv:2609.13377},
  year={2026}
}
\end{document}